\documentclass[preprint,12pt]{elsarticle}

\makeatletter
\def\ps@pprintTitle{%
  \let\@oddhead\@empty
  \let\@evenhead\@empty
  \let\@oddfoot\@empty
  \let\@evenfoot\@empty}
\makeatother
\date{}

\usepackage{comment}
\usepackage{graphicx}
\usepackage{caption}
\usepackage{subcaption}
\usepackage{multirow}
\usepackage{amsmath,amssymb,bm}
\usepackage{algorithm}
\usepackage{algorithmic}
\usepackage{hyperref}
\usepackage{xcolor}
\usepackage{float}
\usepackage[a4paper,margin=2cm]{geometry}

\usepackage{tikz}
\usetikzlibrary{arrows,positioning}
\tikzstyle{block} = [rectangle, rounded corners, draw=black, thick, minimum width=3cm, minimum height=1cm, align=center]
\tikzstyle{arrow} = [->, thick]

\newcommand{\ms}[1]{{\color{blue}#1}}

\journal{International Journal of Heat and Mass Transfer}

\begin{document}

\begin{frontmatter}

\title{A Multi-Fidelity Global Search Framework for Hotspot Prevention in 3D Thermal Design Space}

\author[TUD]{Morteza Sadeghi\corref{cor1}}
\ead{morteza.sadeghi@nmf.tu-darmstadt.de}

\author[ubc]{Hadi Keramati}
\ead{hadi.keramati@ubc.ca}

\author[ncsu]{Sajjad Bigham}
\ead{sbigham@ncsu.edu}

\cortext[cor1]{Corresponding author: morteza.sadeghi@nmf.tu-darmstadt.de}

\address[TUD]{Institute for Nano- and Microfluidics,
Department of Mechanical Engineering, Technical University of Darmstadt, Darmstadt, Germany.}
\address[ubc]{Department of Mechanical Engineering, 
The University of British Columbia, Vancouver, Canada.}

\address[ncsu]{Department of Mechanical and Aerospace Engineering, 
North Carolina State University, Raleigh, NC 27695-7910, USA.}
\begin{abstract}
We present a Bézier-based Multi-Fidelity Thermal Optimization Framework, which is a computationally efficient methodology for the global optimization of 3D heat sinks. The flexible Bézier-parameterized fin geometries and the adopted multi-fidelity pseudo-3D thermal modeling strategy meet at a balance between accuracy and computational cost. In this method, the smooth and compact Bézier representation of fins defines the design space from which diverse topologies can be generated with minimal design variables. A global optimizer, the Covariance Matrix Adaptation Evolution Strategy, minimizes the pressure drop with respect to a given surface-average temperature constraint to achieve improvement in the pressure loss. In the framework, the pseudo-3D model couples two thermally interacting 2D layers: a thermofluid layer representing the fluid domain passing through the fins, and a conductive base plate representing the surface where excessive average temperature is to be avoided. Both layers are coupled with calibrated heat transfer coefficients obtained from high-fidelity 3D simulations. For several fin geometries, the proposed framework has been validated by comparing the pseudo-3D results with those of full 3D simulations, which yielded good agreement in terms of temperature distribution and pressure drops when the computational cost was reduced by several orders of magnitude. Optimization results show that it attains up to 50\% pressure loss reduction compared to conventional straight-fin configurations, and it reveals a clear trade-off between thermal performance and hydraulic efficiency. Thus, the proposed method forms a new basis for fast, geometry-flexible, and optimized heat sink design, enabling efficient exploration of complex geometries.
\begin{comment}
We present an efficient global approach for heat sink optimization using Bézier-based fins combined with a scalable multi-fidelity 3D heat transfer modeling framework. Here, the thermal design space is parameterized using Bézier curves, enabling flexible and smooth geometric representations of fin structures. A global optimization algorithm, covariance-matrix-adaptation evolution strategy (CMA-ES), is employed to minimize pressure drop while enforcing a constraint on the surface-average temperature. To alleviate the high computational costs associated with 3D simulations, a pseudo-3D modeling approach that couples two thermally interacting 2D surfaces: a thermofluid layer representing the fluid domain passing through the fins, and a conductive base plate representing the surface where excessive average temperature is to be avoided. The convective and conductive heat transfer coefficients are calibrated using full 3D simulations and validated across various fin geometries. To validate the results, we also applied extrusion along the fin-height direction to construct 3D geometries. Results show a good agreement between the pseudo-3D and full 3D models in both temperature distribution and pressure drop, while considerably reducing the computational cost. Our results show a pressure loss reduction by up to almost 50\% compared to conventional straight fins, depending on the thermal constraint, thereby revealing a trade-off between thermal performance and pressure losses.  
\end{comment}
\end{abstract}

\end{frontmatter}
\section{Introduction}
Optimization of heat sink geometry is crucial in efficient thermal management of computer processors \cite{husain2008shape}, power electronics \cite{maranzana2004design}, light-emitting diodes (LEDs) \cite{huang2017design}, electric vehicles \cite{hetsroni2002uniform, khan2022design}, aerospace systems \cite{khattak2019air,martinez2019design}, medical devices, \cite{naphon2019ann} renewable energy devices\cite{ref9}, and is a vital component in minimizing the operating temperature of electronic equipments\cite{ref10}. As demands for energy efficiency and environmental awareness grow, the development of effective methods to identify optimal heat sink designs is gaining more attention \ms{\cite{ahmed2018optimization}}. Currently, the design life cycle of thermal management devices is time-consuming, requiring iterative processes of design selection, geometry generation, performance simulation, and simulation-based optimization. 

Over the past few years, various heat sink geometries in combination with different optimization techniques have been examined to generate geometries that can be adaptively modified during optimization through shrinking or expanding the physical size of heat transfer features \cite{wang2011multi,subasi2016multi,chamoli2019numerical,ghasemi2021multi}. In such shape optimization approaches, the design process is tailored to determine the best dimensions within a specified class of heat transfer features \cite{arie2017air}. To accelerate the optimization process, hybrid approaches have been examined that employ nonparametric machine learning models as surrogates for simulation results, thereby enhancing performance evaluation and iteration speed \cite{liu2017multi}, while keeping the geometry and fin design pre-defined. More recently, artificial neural networks (ANNs) have been employed to expedite the design iteration process through simultaneous estimation of thermal and hydraulic performance, while the overall shape of heat transfer features still remained predetermined \cite{safikhani2014multi}. 
In contrast, traditional topology optimization approaches, such as density-based methods \cite{dilgen2018density, hoghoj2020topology}, have been suggested for optimizing the design of heat transfer features. Although these methods have proven effective in identifying optimal structural designs, they often become trapped in local optima due to the highly nonlinear relationship between design variables and thermal performance metrics arising from convection–diffusion effects \cite{dilgen2018density}. Furthermore, the density-based methods, when applied for heat sink optimization, result in closely spaced heat transfer features or disconnected flow pathways \ms{\cite{Lazarov2016}}. To reduce computational time, particularly in 3D problems, hybrid approaches combining density-based optimization with simplified thermal resistance modeling have been employed \cite{haertel2018topology,huang2022pseudo}. To further mitigate the practical challenges in heat sink/exchanger design using gradient-based methods, modeling frameworks employing the Hadamard boundary variation method were introduced for 2D and 3D domains, imposing thickness constraints through a level-set function \cite{feppon2021body,feppon2020topology}. These constraints ensure distinct hot and cold flow domains, preventing leakage across the separating wall. However, prior attempts have shown convergence within fewer than 200 iterations, thereby suggesting that the design space is not adequately explored in shape-derivative approaches, in particular for 3D problems \cite{feppon2020topology}. 

For black-box, multi-modal design spaces with expensive evaluations, global derivative-free optimizers are widely employed \cite{kimura2022statistical,guo2024hybrid,guo2025advancing}. The covariance-matrix-adaptation evolution strategy (CMA-ES), in particular, adapts its search distribution to curved, ill-conditioned landscapes and has been widely used in aerodynamic and thermo-fluid shape optimization algorithms \cite{wang2023topology,park2025optimization,fujii2019topology}. To explore the design space more effectively, several studies have suggested pixel- and voxel-based geometry representations combined with genetic algorithms \cite{mekki2021genetic,mekki2022voxel}. This approach is particularly promising for enforcing manufacturing constraints on fin designs while allowing sufficient design freedom. However, the high dimensionality of voxel-based designs limits scalability. To address the curse of dimensionality in heat sink design, Bézier curves have been considered, parameterizing the heat sink geometry instead of pixels, paired with reinforcement learning and agent-based approaches \cite{keramati2022deep,ravanji2024optimising,keramati2022generative}. While effective, these approaches have been limited to two-dimensional domains, ignoring the temperature variations along the third dimension as required for three-dimensional problems.

The large computational cost of high-fidelity 3D conjugate heat transfer simulations creates a fundamental roadblock to any iterative optimization routine. A single 3D simulation often requires hours or days to compute, while thousands are needed for a global search. This has motivated the recent development of multi-fidelity techniques. Purely data-driven models, such as neural networks, can provide fast predictions but require voluminous training data from high-fidelity simulations and may fail to generalize to new geometries not included in that training. An alternative is to use physics-based, reduced-order models that preserve essential physical coupling but approximate the governing equations somehow. Models of this type—such as the pseudo-3D approach explored in this work—offer a compromise, significantly reducing the computational cost but with a much stronger underpinning of physics than purely data-driven surrogates, and therefore are well-suited to being embedded within an optimization loop.

To address the above shortcoming, we present an approach that combines boundary representation of plate-fin geometries with a multi-fidelity, resistance-based method for optimizing three-dimensional heat transfer problems. A 3D simulation environment is constructed using two interacting surfaces parametrized by the control points of the Bézier curves. This allows the design generation process to be integrated with a global search engine, effectively exploring the design space. The proposed geometry parametrization, coupled with a highly efficient multi-fidelity simulation for identifying hotspots and pressure drop, enables rapid constraint-based optimization, resulting in highly flexible geometries.

\section{Problem Modeling}
\subsection{General Description}
The goal of the present study is to develop an algorithm for the optimization of 3D heat sink geometries. A steady-state, laminar, incompressible fluid flow with conjugate heat transfer is considered. Figure 1a shows a 3D schematic of the heat sink considered in this work. The algorithm optimizes the geometry of heat transfer features to simultaneously minimize the base plate temperature and pressure drop penalty. To reduce the computational time and cost associated with conventional shape optimization techniques, a pseudo 3D model, represented in Fig. 1 b, is employed. The pseudo model approximates the original 3D conjugate heat transfer problem as two 2D thermally coupled problems. 
The pseudo 3D approach consists of a 2D thermofluid layer, $\Omega_f \cup \Omega_d\in \mathbb{R}^2$,  that represents the original 3D thermofluid problem in the fluid flow and heat sink fins and a 2D conductive base plate layer, $\Omega_{bp}\in \mathbb{R}^2$, that represents the original 3D thermal diffusion problem in the heat sink base plate. $\Omega_{f}$ is a non-optimizable pure fluid area outside the design domain and $\Omega_d$ is the design domain area occupied by fins ($\Omega_{d,fin}$) and the incompressible ﬂuid ($\Omega_{d,fl}$). The base plate of the heat sink, $\Omega_{bp}$, is below the design domain and is identical to the area of the design domain in terms of dimensions. It consists of attached-fin ($\Omega_{bp,fin}$) and fin-free ($\Omega_{bp,fl}$) regions. The modeling of the thermal coupling between the base plate and the thermofluid design layer is described in Section \ms{\ref{sec:CFD Environment}}. Therefore, the fluid and heat transfer problems in the thermofluid design layer are modeled two-dimensionally. The 2D assumption is motivated by the fact that the fin height is considerably larger than the xy-dimensions. Also, the original 3D thermal diffusion problem in the base plate is simplified to a 2D problem as the xy dimensions of the base plate are much larger than its height. The validity of these simplifications is assessed in Section \ref{sec:results} (see Figure \ref{fig:error_validation_combined}), and further examined in Section \ref{sec:results}, Table \ref{tab:pseudo_vs_3D}, where the simulation results of the full 3D and pseudo-3D approaches are compared.

The solid parts ($\Omega_{d,fin}$, $\Omega_{bp,fin}$) consist of the identical composite Bézier curves defined by movable control points, which introduce incremental changes to the fin-free domains ($\Omega_{d,fl}$, $\Omega_{bp,fl}$). Defining the geometry using Boundary Representation (BREP) facilitates the mathematical implementation of moving boundary conditions during these incremental modifications of the computational domain.
\begin{figure}[htbp]
    \centering
    \begin{minipage}{0.48\textwidth}
        \centering
        \includegraphics[width=0.9\linewidth]{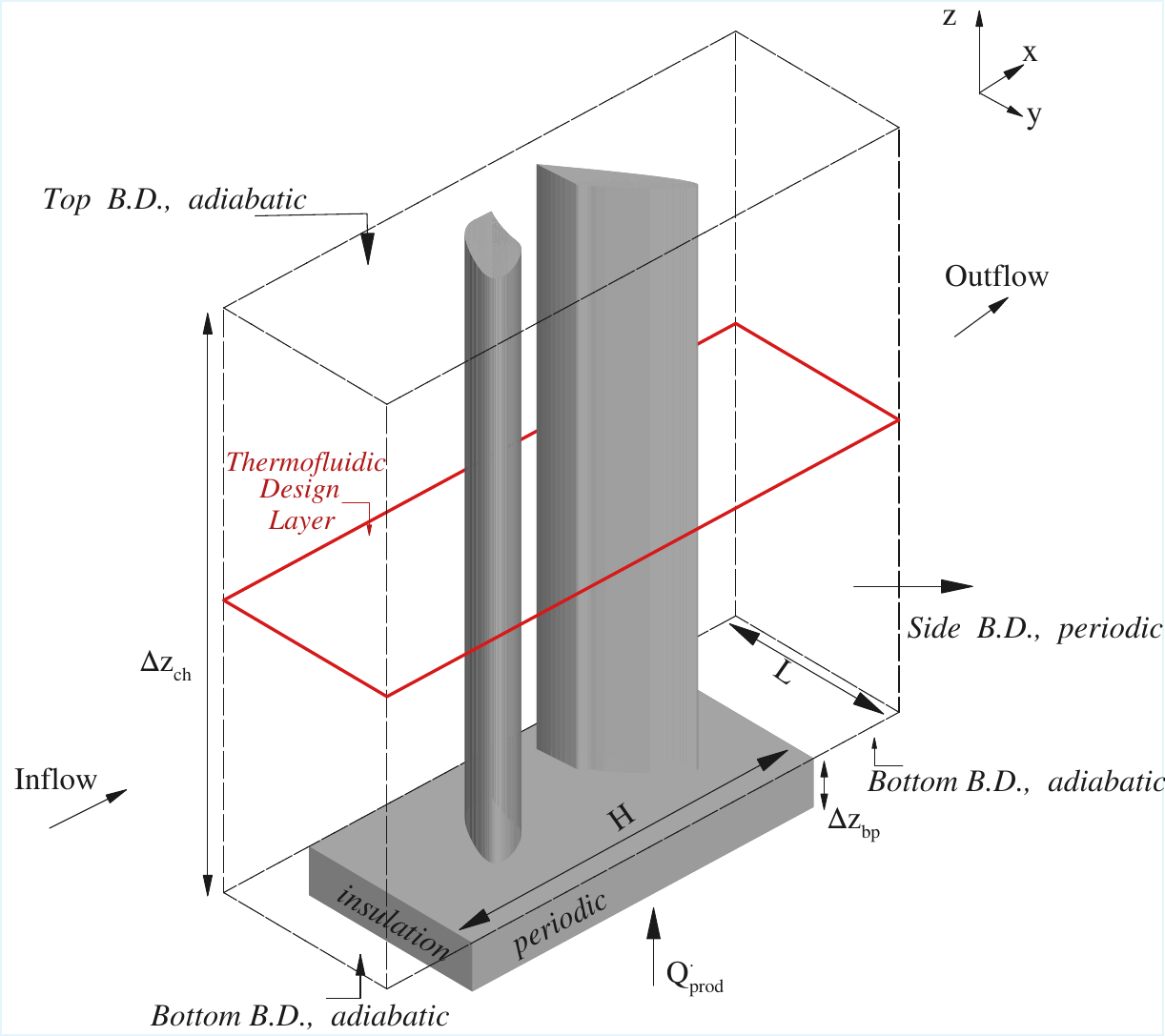}
        \\ (a)
    \end{minipage}\hfill
    \begin{minipage}{0.48\textwidth}
        \centering
        \includegraphics[width=0.9\linewidth]{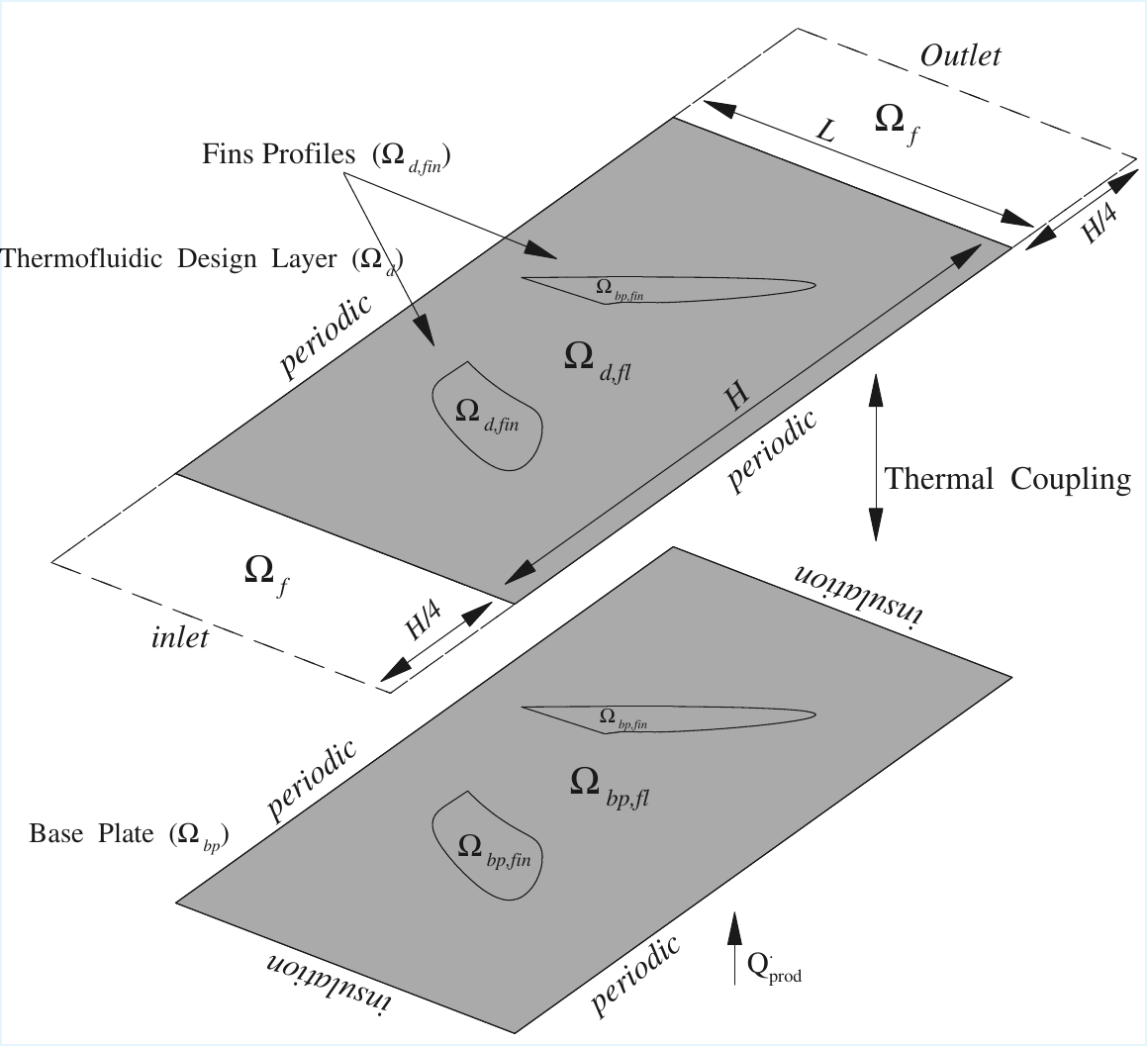}
        \\ (b)
    \end{minipage}
    \caption{%
        (a) 3D schematic view of the problem with boundary conditions. 
        (b) Simplified pseudo-3D model, showing the conversion from the full 3D model to two 2D models, with boundary conditions indicated.
    }
    \label{fig:schematic_models}
\end{figure}
\subsection{Boundary Representation using composite Bézier curves }
In geometric modeling and computer graphics, a composite Bézier curve refers to a curve formed by joining several Bézier curve segments such that the resulting curve is continuous and the segments are connected without gaps. They are mostly used in computer graphics and geometric modeling for image manipulation, and also in Generative Adversarial Network (GAN) for design space exploration. Thanks to the mathematical definition of Bézier curves, which facilitates the implementation of iterative processes, and their continuous derivatives, which are extremely useful for implementing Neumann boundary conditions, we have employed them here to analyze the shape optimization process.
A Bézier curve of degree $n$ is defined by $n+1$ control points, $\mathbf{P}_i$ ($i = 0, 1, 2, \ldots, n$), and the general mathematical form given as follows:
\begin{equation}
\mathbf{B}(t) = \sum_{i=0}^{n} B_{i,n}(t)\,\mathbf{P}_i
\end{equation}
where n is the degree of the curve and $B_{i,n}(t)$ is the i-th Bernstein polynomial defined as:
\begin{equation}
B_{i,n}(t) = \binom{n}{i} (1-t)^{n-i} t^i, \quad \text{where } i = 0, 1, \ldots, n,\; t \in [0,1]
\end{equation}
In our study, we have considered four segments for each fin geometry, and for each segment, we employ a cubic Bézier curve, which simplifies the equation for each segment as follows:
\begin{equation}
\mathbf{B}(t) = (1-t)^3 \mathbf{P}_0 + 3(1-t)^2 t\, \mathbf{P}_1 + 3(1-t)t^2\, \mathbf{P}_2 + t^3 \mathbf{P}_3, \quad t \in [0,1]
\end{equation}
Accordingly, we need 12 control points in total for each fin geometry. A detailed explanation of the closed Bézier shape generation process is provided in Section \ref{sec:Construction of the Closed Bézier Shapes}.

\subsection{Construction of the Closed Bézier Shapes}\label{sec:Construction of the Closed Bézier Shapes}
To construct the closed Bézier curve, the shape is divided into four segments with each segment represented by a cubic Bézier curve. The first step is to define the primary control points of the Bézier segments, which define the endpoints of each segment and are shared between successive curves in order to maintain continuity. Subsequently, the method for generating the intermediate control points, completing the set of four control points for each segment, is described.
\subsubsection{Generating Primary Outline Vertices}
The primary outline vertices represent the $m$ evenly rotationally spaced points defining the initial shape of the polygonal outline before any interpolation. This polygon, formed by connecting the primary control points, serves as the base geometry from which the smoothed Bézier curve is subsequently generated. The $m = 4$ vertices (primary control points) are produced as follows for each fin. First, the initial vertices is evenly distributed at angles, $\theta^{\mathrm{primary}}_i = i \cdot \frac{2\pi}{m}, \quad i = 0, 1, \dots, m-1$ on a circle of radius $r_{\mathrm{max}}$. Second, each vertex's angle is perturbed by $\theta_i = \theta^{\mathrm{primary}}_i + \delta\theta_i \cdot \frac{2\pi}{2m}$. In the third step, a radial deformation of the initial vertices is introduced using the formula $r_i = |\delta r_i| \cdot r_{\mathrm{max}}$, and the polar coordinates of the initial vertices are converted to Cartesian coordinates by $x_i = r_i \cos(\theta_i), \quad y_i = r_i \sin(\theta_i)$. In the next step, the vertices are shifted by the offsets  $(x_{\mathrm{shift}}, y_{\mathrm{shift}})$ to ensure that the polygon can move freely anywhere in the whole area of the design space. 
\subsubsection{Generation of Intermediate Bézier Control Points}\label{sec:Generation of Intermediate}
Once the polygonal base shape is constructed from the primary vertices, each edge of the polygon is then replaced by a cubic Bézier curve. In order to provide smooth transitions, a tangent direction is set to each primary outline vertex by computing a weighted average of incoming and outgoing edge headings as follows:
\begin{equation}
\tilde{\phi}_i = w_i \,\phi_i^{\mathrm{out}} + (1-w_i) \,\phi_{i-1}^{\mathrm{out}} + \mathbf{1}\left(|\phi_i^{\mathrm{out}} - \phi_{i-1}^{\mathrm{out}}| > \pi \right)\pi
\end{equation}
where $\phi_i^{\mathrm{out}}$ is the outgoing edge heading from the initial vertex $P_i$ to $P_{i+1}$, $P_i \to P_{i+1}$, and $\phi_{i-1}^{\mathrm{out}}$ is the incoming edge heading from $P_{i-1}$ to $P_i$, $P_{i-1} \to P_i$, which both are wrapped to $[0, 2\pi)$. Moreover, the weight parameter $w_i$ is evaluated from the curvature control parameter $\eta_i\in[0, 1]$ as below:
\begin{equation}
w_i = 0.5 + 0.5 \, \eta_i \quad\in\quad [0.5, 1]
\end{equation}
and a branch-cut modification term is also added at the end to ensure that averaging takes place along the circle's shorter arc rather than across the 0/2$\pi$ discontinuity. 

For the edge originally defined by the points $P_0 \to P_1$, we relabel the endpoints as $P_0 \to P_3$, since two intermediate control points will be inserted between them. After assigning the tangent directions $\tilde{\phi}_0,  \tilde{\phi}_3$ to $P_0$ and $P_3$, respectively, the chord length between the two consecutive vertices, $d = \|P_3 - P_0\|$, is evaluated. The interior control points $P_1$ and $P_2$ are placed along the tangent directions at the endpoints using a distance $r_m \;=\; 0.707\cdot d \cdot\; r_{\mathrm{mid}}$ where $r_{\mathrm{mid}} \in [0,1]$ is a dimensionless curvature scaling factor. Here, $r_{\mathrm{mid}} \in [0, 1]$ is a curvature scaling parameter, and the constant $0.707 = \sqrt{2}/2$ is taken from the circular-arc approximation, where placing the intermediate control points at $\sqrt{2}/2$ of the chord length along the tangent lines leads to a very close match to the true arc. Geometrically, $r_m$ performs as an effective arc radius that controls the bulge extent between two primary vertices $P_0$ and $P_3$; larger $r_m$ produces a more pronounced curvature, smaller $r_m$ produces a flatter segment.
When $r_{\mathrm{mid}}=1$, the Bézier segment very closely resembles the circular quarter-arc placement.

Given the endpoint tangents and coordinates, the location of the intermediate control points can be found via
$P_1 = P_0 + r_m \, (\cos\tilde{\phi}_0,\, \sin\tilde{\phi}_0),
P_2 =\; P_3 + r_m \, \big(\cos(\tilde{\phi}_3+\pi),\, \sin(\tilde{\phi}_3+\pi)\big)
$. $P_1$ lies along the forward tangent direction at $P_0$ while $P_2$ lines along the backward tangent direction at $P_3$ (+$\pi$ rotation with respect to $\tilde{\phi}_3$). By repeating this procedure for each segment of the outline, a sequence of Bézier curves is created that join smoothly to form a closed, smooth contour for each fin.

\subsection{CFD Environment}\label{sec:CFD Environment} 
The CFD environment created for our pseudo 3D approach is shown in Fig. 1b. For both the simplified 2D model and the corresponding 3D validation model, a thickness of the base plate $\Delta{z}_{bp}=H/8$, a channel height $\Delta{z}_{ch}=1.5H$, and a base plate width $L=0.5H$, with $H=10 \text{mm}$ are assumed.
Assuming an incompressible fluid and two-dimensional flow in the xy-plane, the continuity equation and Navier-Stokes equation are as follows:

\begin{equation}
\rho_f\, \nabla \cdot \mathbf{u} = 0 
\label{eq:continuity}
\end{equation}
\begin{equation}
\rho_f (\mathbf{u} \cdot \nabla) \mathbf{u} = \nabla \cdot \left[-p\mathbf{I} + \mathbf{K}\right]
\label{eq:momentum}
\end{equation}
\begin{equation}
\mathbf{K} = \mu_f \left( \nabla \mathbf{u} + (\nabla \mathbf{u})^{\mathrm{T}} \right)
\label{eq:momentum_2}
\end{equation}

where $\rho_f$ is the fluid density, $\mathbf{u}$ is the fluid velocity vector, $p$ is the pressure field, and $\mu_f$ is the dynamic viscosity of the fluid.
A fluid velocity of $u_{in}=1(\text{m/s})$ is prescribed at the inlet, and the pressure outlet boundary condition, $p=0$, is applied at the outlet. Moreover, periodic flow conditions are imposed on the side boundaries to reflect the periodic feature of the heat sink in the direction transverse to the airflow. To evaluate pressure drop, the following formula is used:
\begin{equation}
\Delta \overline{p} =
\overline{p}_{inlet}
-
\overline{p}_{outlet}
\label{eq:pressure_difference}
\end{equation}
where
\begin{equation}
\overline{p}_{\Gamma} =
\frac{\displaystyle\int_{\Gamma} p\,d\Xi}{\displaystyle\int_{\Gamma} d\Xi}
\end{equation}
where \( d\Xi \) denotes a line element (\( dy \)) or an area element (\( dA \)), depending on the dimension of \( \Gamma \).

In the thermofluid layer outside the design domain ($\Omega_f$), the 2D thermal convection-diffusion equation without heat source or heat sink is solved, which is given by:
\begin{equation}
\rho_f \, c_f \, \mathbf{u} \cdot \nabla T - \nabla \cdot (k_f \nabla T) = 0 \quad \text{in } \Omega_f
\label{eq:energy1}
\end{equation}

where $T$ is the temperature field in the thermofluid layer, $c_f$ the specific fluid heat capacity, and $k_f$ the thermal conductivity of the fluid. Within the design domain ($\Omega_{d}$), the following 2D thermal convection-diffusion equations are solved in the solid ($\Omega_{d,fin}$) and fluid parts ($\Omega_{d,fl}$), respectively:
\begin{equation}
- \nabla \cdot \left( k_s \, \nabla T \right) 
= \frac{\dot{q}_{\text{inter,s}}}{\Delta z_{ch}}
\quad \text{in } \Omega_{d,fin}
\end{equation}
\begin{equation}
\rho_f \, c_f \, \mathbf{u} \cdot \nabla T - \nabla \cdot (k_f \nabla T)
= \frac{\dot{q}_{\text{inter,f}}}{\Delta z_{ch}}
\quad \text{in } \Omega_{d,fl}
\end{equation}
The thermal coupling between the two plates arises from heat transfer from the solid base plate to both the solid and fluid regions of the thermofluid design layer. These interactions are modeled as $\dot{q}_{\text{inter,s}} = h_s (T_{bp}-T)$, and $\dot{q}_{\text{inter,f}} = h_f (T_{bp}-T)$, respectively, where $T_{bp}$ represents the temperature distribution in the solid base plate. Here, $h_f$ describes the convective heat transfer coefficient between the plate and the fluid and $h_s$ represents the conductive heat transfer in the z-direction inside the fins which is evaluated by the method explained in the next section. The following 2D heat conduction problems is solved in the solid base plate ($\Omega_{bp}$), including both the fin-attached ($\Omega_{bp,fin}$) and fin-free regions ($\Omega_{bp,fl}$) :
\begin{equation}
-\nabla \cdot \left( k_s \nabla T_{bp} \right) = \frac{\dot{Q}_{prod}}{V_{bp}} - \frac{\dot{q}_{\text{inter}}}{\Delta z_{bp}}
\quad \text{in}~\Omega_{bp}
\label{eq:energy2}
\end{equation}
where $k_s$ is the base plate thermal conductivity, $T_{bp}$ is the temperature field in the base plate, $\dot{Q}_ {prod}$ is the prescribed heat rate production in the base plate, $V_{bp}$ is the volume of the base plate, and $\Delta z_{bp}$ is the height of the base plate. Throughout this study, we assume a uniform volumetric heat generation rate $\Dot Q_ {prod}$, within the solid base plate. The term $\dot{q}_{\text{inter}}$ is equal to $\dot{q}_{\text{inter,s}}=h_s (T_{bp}-T)$ for the area of the solid base plate covered by the fins ($\Omega_{bp,fin}$), and is equal to $\dot{q}_{\text{inter,f}}=h_f (T_{bp}-T)$ in the area having contact to the fluid flow ($\Omega_{bp,fl}$), as shown in Fig.~\ref{fig:schematic_models}b  .

The fluid temperature at inlet is fixed at $T_{in}=298.15 \text{K}$ and the outflow boundary condition, $n.\nabla{T}=0$, is applied the outlet. Periodic boundary conditions are applied at the side boundaries to reflect the periodicity of the heat sink. For the base plate, insulation (adiabatic) boundary conditions are imposed on the front and back sides. To evaluate average surface temperature, the following formula is used:
\begin{equation}
\overline{T}_{\Gamma} =
\frac{\displaystyle\int_{\Gamma} T\,d\Xi}{\displaystyle\int_{\Gamma} d\Xi}
\label{eq:Tavg}
\end{equation}
where \( d\Xi \) denotes a line element (\( dy \)) or an area element (\( dA \)), depending on the dimension of \( \Gamma \).

\subsection{Determination of  $h_f$ and $h_s$}\label{sec:determination_h}
The accuracy of the temperature evaluation depends on the precise specification of the two parameters, $h_f$ and $h_s$. These parameters govern the heat transfer from the solid base plate to the fluid flowing over the chip and to the attached solid fins, respectively. Specifically, $h_f$ represents the convective heat transfer between the base plate and the airflow above it, while $h_s$ represents the conductive heat transfer coefficient between the base plate and the attached fins. We determine these coefficients by averaging the results of a set of full 3D simulations performed for different fin configurations. The expressions used to evaluate the heat transfer coefficients are provided below. 

\begin{equation}
h_{f} = \frac{\dot{q}_{\Omega_{bp,fl}}}{A_{\Omega_{bp,fl}} \left( \overline{T}_{\Omega_{bp,fl}} - \overline{T}_{fl} \right)}
\end{equation}
where $\dot{q}_{\Omega_{bp,fl}}$ is the heat transfer rate from the fin-free area of the solid base plate ($\Omega_{bp,fl}$) to the fluid, $A_{\Omega_{bp,fl}}$ is the fin-free area of the solid base plate, $\overline{T}_{\Omega_{bp,fl}}$ is the average temperature of the fin-free surface, and $\overline{T}_{fl}$ is the volume weighted average temperature of fluid in the channel domain. $h_s$ reflects the conduction heat transfer between the
base plate surface and the fins, and accordingly is called the nominal convection heat transfer coefficient. This coefficient can be evaluated like $h_f$ as follows:
\begin{equation}
h_{s} = \frac{\dot{q}_{\Omega_{bp,fin}}}{A_{\Omega_{bp,fin}} \left( \overline{T}_{\Omega_{bp,fin}} - \overline{T}_{fin} \right)}
\end{equation}
where $\dot{q}_{\Omega_{bp,fin}}$ is the heat transfer rate from the attached-fin area of the solid base plate ($\Omega_{bp,fin}$) to the fluid, 
$A_{\Omega_{bp,fin}}$ is the attached-fin area of the solid base plate, $\overline{T}_{\Omega_{bp,fin}}$ is the average temperature of the attached-fin surface, and $\overline{T}_{fin}$ is the volume weighted average temperature of fins. After evaluation of these parameters for different configurations, the average values $h_f=80 (\text{W}/\text{m}^2\text{K})$ and $h_s=44500(\text{W}/\text{m}^2\text{K})$ are obtained.
The accuracy of these values is assessed by applying them to several heat sink designs and comparing the resulting temperature distribution and average base-plate temperature, as discussed in detail in Section \ref{sec:results}.

\subsection{Optimization scheme}
The heat sink optimization problem considered here is a dual-objective optimization problem. The first objective is to reduce the average surface temperature of the base plate, and the second objective is to minimize the pressure loss of the driven fluid. In this study, they are handled as a combined minimization and constraint problem:

\begin{align}
\text{Minimize:} \quad & \Delta \overline{p} \label{eq:optimization}\\
\text{Subject to:} \quad &  \overline{T}_{bp} \leq \overline{T}_{\mathrm{cons}} \nonumber\\
& \text{Continuity equation:} \quad \text{Eq.~(\ref{eq:continuity})} \nonumber\\
& \text{N-S equation:} \quad \text{Eq.~(\ref{eq:momentum}, \ref{eq:momentum_2})} \nonumber\\
& \text{Energy equations:} \quad \text{Eqs.~(\ref{eq:energy1}) - (\ref{eq:energy2})} \nonumber
\end{align}
where $\Delta\overline{p}$ is the pressure drop through the heat sink and $\overline{T}_{cons}$ is the temperature constraint value. Optimization is performed in several different bottom surface average temperature constraints, which results in different designs for each case. 

\section{Optimization Methodology}
\subsection{Decision Vector Structure}\label{sec:decision-vector}
A decision vector X encodes the geometry and placement of fins, and has a length $\text{dim}(X)=2n_f+3mn_f$ where $n_f$ is the number of fins (here $n_f=2$) and $m$ is the number of primary outline vertices per fin (here $m=4$). The first $2n_f$ elements of X define the positional shifts of each fin by moving the centroid of the polygon formed by the primary outline vertices, $
X[0 : n_f) \rightarrow x\text{-shifts}, \quad 
X[n_f : 2n_f) \rightarrow y\text{-shifts}.$ The remaining $3 m n_f$ elements define shape deformations, grouped into triplets for each vertex for all $n_f$ fins. In fact X has a form as follows:
\begin{equation}
X =
\underbrace{[x_{1}, \dots, x_{n_f}, \; y_{1}, \dots, y_{n_f}]}_{\text{Fin position shifts}}
\;\cup\;
\underbrace{[\delta r, \delta \theta, \eta]_{1,1}, \dots, [\delta r, \delta \theta, \eta]_{m, n_f}}_{\text{Fin shape deformation}}\label{eq:design_variables}
\end{equation}
where radial deformation $\delta r_i$ determines the vertices’ distance from the centroid, angular offset, $\delta\theta_i$, introduces a perturbation rotationally to the vertices location, and curvature control, $\eta_i$, influences the smoothness of the connection to adjacent vertices. Based on this set of parameters, the geometry is constructed according to the method outlined in Section \ref{sec:Construction of the Closed Bézier Shapes}.

\subsection{Cost Function and Penalties}\label{sec:costfunction}
In our study the cost function is:
\begin{equation}
J(X) = -\Delta p + P_{\mathrm{geom}}(X) + P_{\mathrm{thermal}}(X) \label{penalty}   
\end{equation}
where $\Delta p$ is the pressure drop through the heat sink evaluated from Eq.~(\ref{eq:pressure_difference}) and $P_{\mathrm{geom}}$ and $P_\mathrm{thermal}$ are the geometric and thermal penalties, respectively. The geometric penalty can be calculated from the following formula:
\begin{equation}
\resizebox{0.95\hsize}{!}{$
P_{\mathrm{geom}}(X) =
\lambda_{\mathrm{geom}} \sum_{p} \Big[
\max(0, x_{\min} - x_{\min{geom}}) +
\max(0, x_p - x_{\max{geom}}) +
\max(0, y_{\min{geom}} - y_p) +
\max(0, y_p - y_{\max{geom}})
\Big]
$}
\end{equation}
enforcing that geometry remains inside the design domain. $x_{\max{geom}}$, $x_{\min{geom}}$ denote the maximum and minimum x-coordinate of the closed Bézier geometry, while $y_{\max{geom}}$, $y_{\min{geom}}$ show the corresponding extrema in the y direction. The quantities $x_\mathrm{min}$, $y_\mathrm{min}$, $x_\mathrm{max}$, $y_\mathrm{max}$ define the rectangular outline of the design domain and $\lambda_\mathrm{geom}$ is a scaling coefficient controlling the geometric penalty weight. 
\\The thermal penalty function enforces that the average base plate temperature remains below the certain value $\overline{T}_\mathrm{cons}$ and can be evaluated as follows:
\begin{equation}
    P_\mathrm{thermal}(X) = 
    \begin{cases}
    \lambda_\mathrm{thermal} (\overline{T}_{bp} - \overline{T}_{\mathrm{cons}}), & \overline{T}_{bp} > \overline{T}_{\mathrm{cons}}, \\
    0, & \text{otherwise}.
    \end{cases}
\end{equation}
where $\lambda_\mathrm{thermal}$ is a scaling coefficient
controlling the thermal penalty weight.

\subsection{Implementation and Investigation of the Optimization Algorithm}
\begin{figure}[h!]
\centering
\begin{tikzpicture}[node distance=2.0cm]

% Style for all blocks
\tikzstyle{block} = [rectangle, rounded corners, draw=black, thick,
                     minimum width=6.5cm, minimum height=1.6cm,
                     align=center, font=\normalsize]

% --- Nodes ---
\node[block, fill=blue!15] (init) {Step 1: Initialization \\[2pt]
    $m^{(0)}, \;\sigma^{(0)}, \;C^{(0)}$};

\node[block, fill=green!15, below of=init] (sample) {Step 2: Sampling \\[2pt]
    Candidates $X_{k}^{(g)}$ (in the form of Eq. (\ref{eq:design_variables}))

   \\[2pt]
    $X_{k}^{(g)} \sim \mathcal{N}(m^{(g)}, \sigma^{(g)^2} C^{(g)})$};

\node[block, fill=white!15, below of=sample, yshift=-0.8cm] (bezier) {Step 3: Building a $\lambda$ population of Bézier curves using $X_{k}^{(g)}$ to produce generation $g$ \\[6pt]
    \includegraphics[width=0.98\linewidth]{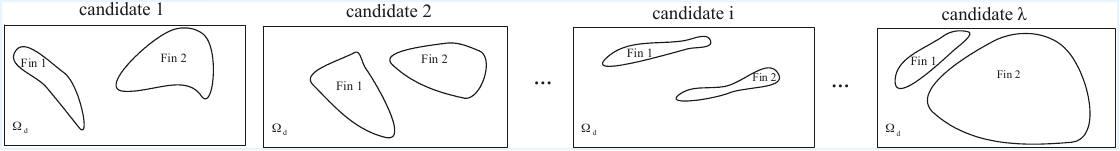}};

\node[block, fill=yellow!25, below of=bezier, yshift=-0.8cm] (eval) {Step 4: Evaluation \\[2pt]
    Cost function $J(X)$ using Eq. (\ref{penalty})
 };

\node[block, fill=red!20, below of=eval] (select) {Step 5: Selection \& Ranking \\[2pt]
    Top $\mu$ candidates};

\node[block, fill=purple!20, below of=select,yshift=-0.20cm] (update) {Step 6: Update Distribution \\[2pt]
    $m^{(g+1)}, \;C^{(g+1)}, \;\sigma^{(g+1)}$};

\node[block, fill=orange!25, below of=select, minimum width=3cm, minimum height=1.2cm ,
xshift=-6cm, yshift=0.9cm] (save) 
    {Step 5.5: \\Save the best candidate};
%\node[block, fill=gray!20, below of=eval, yshift=-2.2cm] (loop) {Repeat until convergence};
\node[block, fill=cyan!20, below of=update] (final) {End: output:optimized domain - Save the best candidate};

% --- Arrows ---
\draw[->, thick] (init) -- (sample);
\draw[->, thick] (sample) -- (bezier);
\draw[->, thick] (bezier) -- (eval);
\draw[->, thick] (eval) -- (select);
% Arrow from Step 5 to Step 6 (main flow)
\draw[->, thick] (select) -- (update) coordinate[midway] (midselup);
\draw[->, thick] 
  (update.east) -- ++(6.15,0)  node[midway, above] {Repeat either until convergence} node[midway, below] {or meeting number of generations} % label above the first horizontal           % go right from Step 6 edge
  |- ([xshift=1.6cm]sample.east)        % go up until aligned with Step 2
  -- (sample.east);                     % go left to Step 2 edge
% Arrow from the midpoint of Step 5→6 arrow to the block's right edge
\draw[->, thick] (midselup) -- (save.east);
\draw[->, thick] (update) -- (final);
%\draw[->, thick] (eval) -- (loop);
%\draw[->, thick] (loop) -- (sample);

\end{tikzpicture}
\caption{Workflow of the CMA-ES algorithm implemented to our problem. The design variables $X_k^{(g)}$ are turned into geometric representations by mapping them into Bézier-curves variations.}
\label{workflow}
\end{figure}

\begin{figure}[H] % Requires \usepackage{float}
    \centering
    \includegraphics[width=1.0\textwidth]{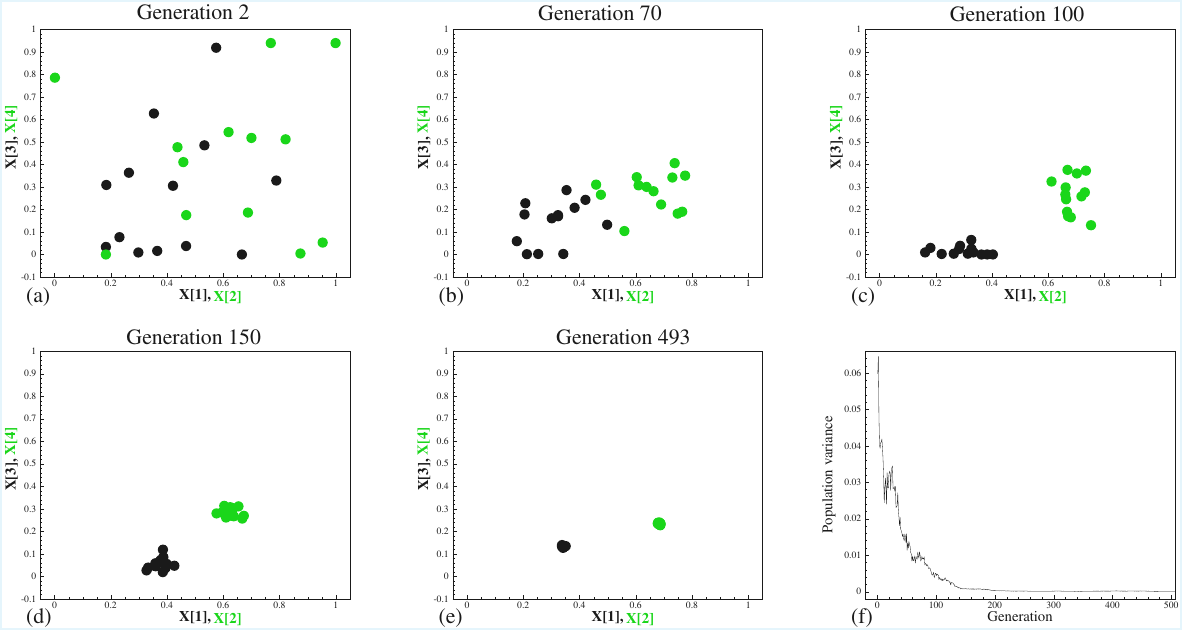}
    \caption{Adaptation of the ellipsoid-shaped search distribution based on the covariance matrix and step size. }
    \label{fig:covariance}
\end{figure}

Among the more recent evolutionary strategies, the Covariance Matrix Adaptation Evolution Strategy (CMA-ES) stands out as an excellent-performing, self-adaptive algorithm with minimal need for manual tuning by the user. Benchmarking of several
algorithms has proved that CMA-ES works well on a large variety
of problems and applications and is one of the most efficient evolutionary strategies for dealing with difficult numerical optimization problems \cite{belaqziz2014irrigation}. Introduced by Hansen and Ostermeier \cite{hansen2023,hansen1996adapting,hansen1997convergence,hansen2001completely}, CMA-ES is a self-adaptive, population-based algorithm which, unlike many traditional point-to-point approaches, operates by moving the population—in the form of a multivariate normal distribution—around the search space.
By means of a self-adaptive exploration procedure, CMA-ES updates the mean and covariance matrix during the optimization process and generates new search regions using a multivariate normal distribution to find the optima of the problem. One of the most important features of this algorithm is its ability to learn the correlations between the parameters and use this information in the optimization process to converge more quickly to the problem's optimum. The effectiveness of this algorithm has been demonstrated across a wide variety of problems and applications \cite{fateen2012evaluation,iruthayarajan2010covariance,ghosh2012differential}.
Evolution strategies can be categorized into three main goroups \((1+1)\)-ES, \((\lambda, \mu)\)-ES, and \((\lambda + \mu)\)-ES. In \((1+1)\)-ES, a single candidate solution produces one offspring per iteration, 
and the better of the two survives. In \((\lambda + \mu)\)-ES, the next generation of \(\mu\) parents is chosen 
from the union of the original \(\mu\) parents and the generated \(\lambda\) offspring. 
In contrast, \((\lambda, \mu)\)-ES chooses the \(\mu\) best points 
only from the \(\lambda\) newly generated offspring, without considering 
the previous individuals. The Covariance Matrix Adaptation Evolution Strategy (CMA-ES) 
belongs to the \((\lambda, \mu)\)-ES category. 
It produces \(\lambda\) candidates by sampling from a 
multivariate \(n\)-dimensional normal distribution, and then selects 
the best \(\mu\) individuals (based on objective function values) 
to continue the search in the next iteration. For the sake of brevity, the physical interpretation of the CMA-ES algorithm implemented for our problem, along with some pertinent formulas, is discussed here, and for more detailed equations, the readers are referred to \cite{hansen2023}.  
In CMA-ES, candidates are selected from a multivariate normal distribution
$
x_k^{(g)} \sim \mathcal{N}\big(m^{(g)}, \, \sigma^{(g)^2} C^{(g)}\big)$ where $m^{(g)}$ is the mean vector, $\sigma^{(g)}$ is the global step-size, 
and $C^{(g)}$ is the covariance matrix at generation $g$. 
The covariance matrix $C^{(g)}$ specifies the shape and orientation 
of the candidate's distribution. In fact, the multivariate normal distribution geometrically corresponds to an ellipsoid in the search space, $\mathcal{E} = \left\{ x \in \mathbb{R}^n \;\middle|\; 
\big(x - m^{(g)}\big)^T \, {C^{(g)}}^{-1} \, \big(x - m^{(g)}\big) = 1 \right\}$, centered at $m^{(g)}$, with principal axes aligned with the eigenvectors of $C^{(g)}$,
and diameters equal to $d_i = 2 \, \sigma^{(g)} \, \sqrt{\lambda_i}\, (i = 1, \dots n$) along each principal axis. Here, $\sigma^{(g)}$ shows the global step-size at generation $g$, and $\lambda_i$ is the $i$-the eigenvalue of the covariance matrix $C^{(g)}$. Accordingly, the orientation of the ellipsoid, reflecting the directions of correlation between the parameters (principal axes), is given by the eigenvectors of $C^{(g)}$ and the axis lengths are determined by the eigenvalues of $C^{(g)}$. The ellipsoid has long diameters or is stretched more in the directions of large eigenvalues (more exploration in that direction), while it has short diameters or is contracted in the irrelevant directions of small eigenvalues. It is also obvious that if all eigenvalues are equal, there will be no preferred direction, and the ellipsoid changes to a sphere. $\sigma^{(g)}$ is the global step-size and is in fact a scaling factor, which a large value for corresponds to a large ellipsoid and an expanded exploration area, while a small value results in a tight ellipsoid and a locally refined search region. \\ At the first step, shown in Fig. (\ref{workflow}), the distribution of the candidates is initialized around the mean $m^{(g)}$, and is equally spread in all directions ($C^{(0)}=I_n$) with an initial radius controlled by $\sigma^{(0)}>0$. At this step, since there is no knowledge about the optimized path, the search begins isotropically in all directions in a spherical cloud. In the next step, offspring are sampled in an ellipsoid-shaped cloud centered at $m^{(g)}$ with orientation controlled by $C^{(g)}$ and size scaled by $\sigma^{(g)}$, using the formula $\bm{X}_k^{(g)} = m^{(g)} + \sigma^{(g)} \cdot \left( C^{(g)} \right)^{1/2} \bm{z}_k^{(g)}$ where $\bm{z}_k^{(g)} \sim \mathcal{N}(0, I)$ stands for the normal distribution. Each $\bm{X}_k^{(g)}$ is a decision vector representing the design variables shown in Eq. (\ref{eq:design_variables}). In step 3, each sampled vector $\bm{X}_k^{(g)}$ is mapped into a geometric representation of the domain using the procedure explained in detail in section \ref{sec:Construction of the Closed Bézier Shapes}. After building all geometries of generation $g$, each design is evaluated using the cost function given in section \ref{sec:costfunction}. The offspring are ranked by the cost function value, and the best $\mu$ individuals are selected as $\bm{x}_{i:\lambda}^{(g)}$, $i=1,\ldots,\mu$ in step 5.   Moreover, to ensure global progress, we have also added a block in step 5.5 to store the best offspring found so far. The core feature of the self-adaptation mechanism of CMA-ES is in step 6, where all parameters controlling the evolution of the search area are updated altogether. In this step, the new mean is a weighted average of the best $\mu$ 
candidates found in the previous generation, evaluated by the formula $m^{(g+1)} = \sum_{i=1}^{\mu} w_i \bm{x}_{i:\lambda}^{(g)}$. 
Physically this means that the center of the ellipsoid cloud is pulled towards the best-performing designs by placing heavier weights on fitter candidates. The shape and the overall scale of the search ellipsoid is adapted by updating the covariance matrix ($C^{(g+1)}$) and step-size ($\sigma^{(g+1)}$), respectively. It means that if several good offspring lie along a diagonal, the ellipsoid stretches and rotates to align with that diagonal. In contrast, directions with little success shrink. Moreover, if the progress is steady and consistent, bigger steps are taken, while a stalled or oscillating situation leads to a smaller step size. Near the optimum solution, to provide a fine-grained convergence $\sigma^{(g)}$ decreases. This could be imagined like learning the way to a valley of a landscape and adjusting your search pattern to follow it. Implementation of this update procedure for both the covariance matrix and the step size is performed by using the evolution path and the vector differences between the $\mu$ best individuals from the current and previous generations, which accumulates directional information over multiple generations. This procedure from step 2 to step 6 iterates over and over to get a budget-based termination after a fixed number of generations or to get a convergence-based termination based on the shrinkage of the covariance ellipsoid or improvement of the function values. At the final stage, the best overall value is saved and considered as the optimized solution. It is also worth mentioning that in order to evaluate the cost function, we employed an automated workflow integrating Python, MATLAB, and COMSOL. The design parameter sets are first generated in the Python environment and then passed to MATLAB. Through the COMSOL–MATLAB programming interface, the corresponding \texttt{.mph} model is constructed and simulated within COMSOL. The simulation results are then returned to MATLAB and subsequently transferred back to Python for optimization. \\
An illustration of the mechanism how the CMA-ES converges to the optimal solution of our problem is presented in Fig. (\ref{fig:covariance}). The variables $X[1]$ and $X[3]$ define the positioning of fin 1, while $X[2]$ and $X[4]$ correspond to the positioning of fin 2 as shown in Eq. (\ref{eq:design_variables}). As represented in part (a), at the early stages, the population is widely spread across the entire search area and the sampling distribution area is mostly like a sphere instead of an ellipsoid. This is due to the fact that the covariance matrix is almost close to unity, $C^{(2)}=I_n$, meaning that there is no preferred direction in the sampling at this stage. As expected, in this stage, the algorithm is broadly exploring without focusing on a specific region. In part (b), the ellipsoid search area starts elongation along a preferred search direction where some improvements have been observed. This shows that the algorithm has started learning the correlation between the variables, adapting to the slope of the cost function values. It is no longer like a sphere but an ellipsoid tilted to align with that preferred direction. In part (c), the population distribution area is more concentrated, the covariance ellipsoid ($C^{(100)}$) has shrunk in scale, and the step size ($\sigma^{(100)}$) is reduced. This leads to a narrower and better-aligned improvement direction. The successful directions from the earlier offspring are reinforced, while the unsuccessful ones are suppressed, reducing the exploration phase. In part (d), the covariance ellipsoid continues shrinking further, leading to a tighter clustering. The variance in the irrelevant direction further reduces, which shows that the algorithm intensifies searching near the more promising area and has moved toward the exploitation phase. In part (e), the ellipsoid is extremely tight and collapsed around the optimum area, and almost all the variance is significantly decayed except in the most relevant principal axes. An overview of the population variance changes is shown in part (f). It shows that the overall variance of the population decreases monotonically with some oscillations. Initially, when the algorithm is in the exploration phase, the variance is large; however, over time, the variance decreases in the exploitation phase. In this phase, the offspring collapses to the optimum area. Moreover, oscillations correspond to CMA-ES’s adaptive mechanism, where the step-size increases slightly when exploration is needed and decreases near the optimum.   

\clearpage

\section{Results and Discussions}\label{sec:results}
First, the performance of the pseudo-3D approach is validated against full 3D simulations, focusing on the velocity field, temperature distribution, average surface temperature, and pressure drop penalty across different fin geometries. The comparison is carried out under steady-state conditions with air as the coolant and aluminum as the solid domain. The geometrical and physical constants considered in this study are summarized in Table \ref{tab:constants}.
\begin{table}[h!]
\centering
\caption{Physical properties and boundary conditions used in the simulations.}
\begin{tabular}{llc}
\hline
\textbf{Parameter} & \textbf{Description} & \textbf{Value} \\
\hline
$c_{f}$   & Fluid specific heat capacity      & $1006 \; \text{J/(kg·K)}$ \\
$\mu_{f}$   & Fluid dynamic viscosity           & $1.94 \times 10^{-5} \; \text{Pa·s}$ \\
$\rho_{f}$  & Fluid density                     & $1.204 \; \text{kg/m}^3$ \\
$k_{f}$     & Fluid thermal conductivity        & $0.024 \; \text{W/(m·K)}$ \\
$\rho_{s}$  & Solid density (aluminum)          & $2700 \; \text{kg/m}^3$ \\
$k_{s}$     & Solid thermal conductivity        & $237 \; \text{W/(m·K)}$ \\
$c_{s}$   & Solid specific heat capacity      & $900 \; \text{J/(kg·K)}$ \\
$T_{\text{inlet}}$ & Inlet temperature          & $293.15 \; \text{K}$ \\
$u_{\text{inlet}}$ & Inlet velocity             & $1 \; \text{m/s}$ \\
$\dot{Q}_{prod}$       & Applied surface heat flux         & $1 \times 10^{5} \; \text{W/m}^2$ \\
\hline
$L$         & Base-plate width             & $0.5H$ \\
$\Delta z_{bp}$ & Base plate thickness & $H/8$ \\
$\Delta z_{\text{chan}}$ & Channel height        & $1.5H$ \\
\hline
\label{tab:constants}
\end{tabular}
\end{table}

\begin{figure}[H]
    \centering

    % Top row: (a) and (b)
    \begin{subfigure}[b]{0.48\textwidth}
        \includegraphics[width=\textwidth]{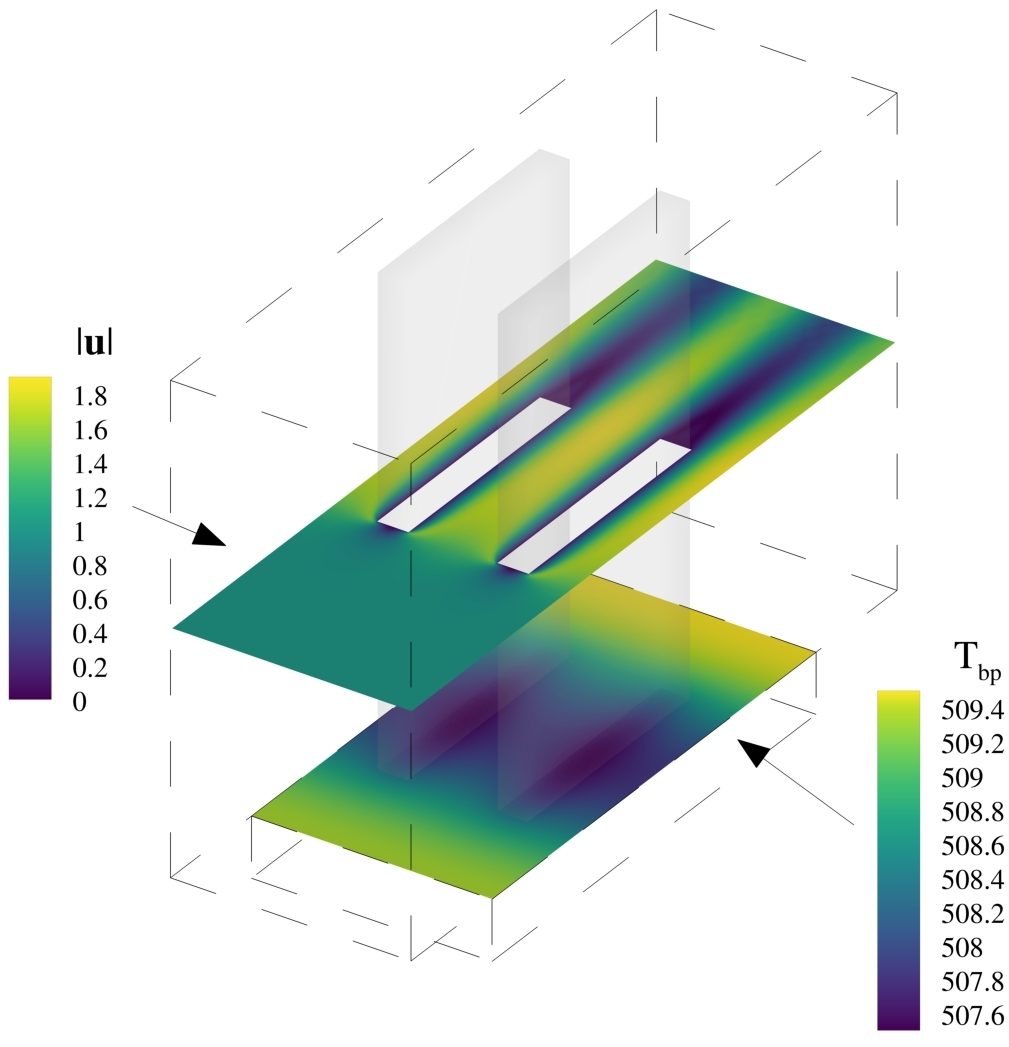}
        \captionsetup{labelformat=empty}
        \caption{(a)}
    \end{subfigure}
    \hfill
       \begin{subfigure}[b]{0.48\textwidth}
        \includegraphics[width=\textwidth]{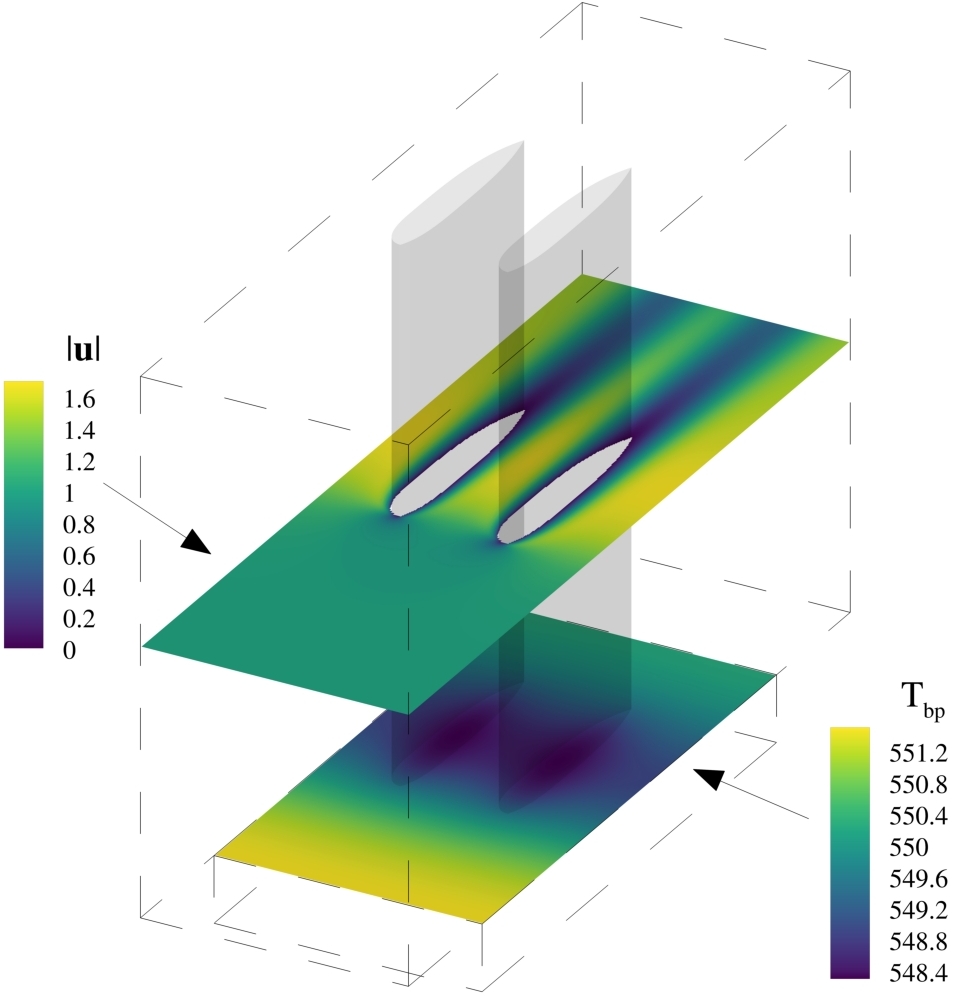}
        \captionsetup{labelformat=empty}
        \caption{(b)}
    \end{subfigure} 

    %    \begin{subfigure}[b]{0.45\textwidth}
%        \includegraphics[width=\textwidth]%{diff_T_number_1_cropped.pdf}
%        \captionsetup{labelformat=empty}
%        \caption{(b)}
%    \end{subfigure}

    \vspace{1em}  % Optional vertical spacing
           \begin{subfigure}[b]{0.48\textwidth}
        \includegraphics[width=\textwidth]{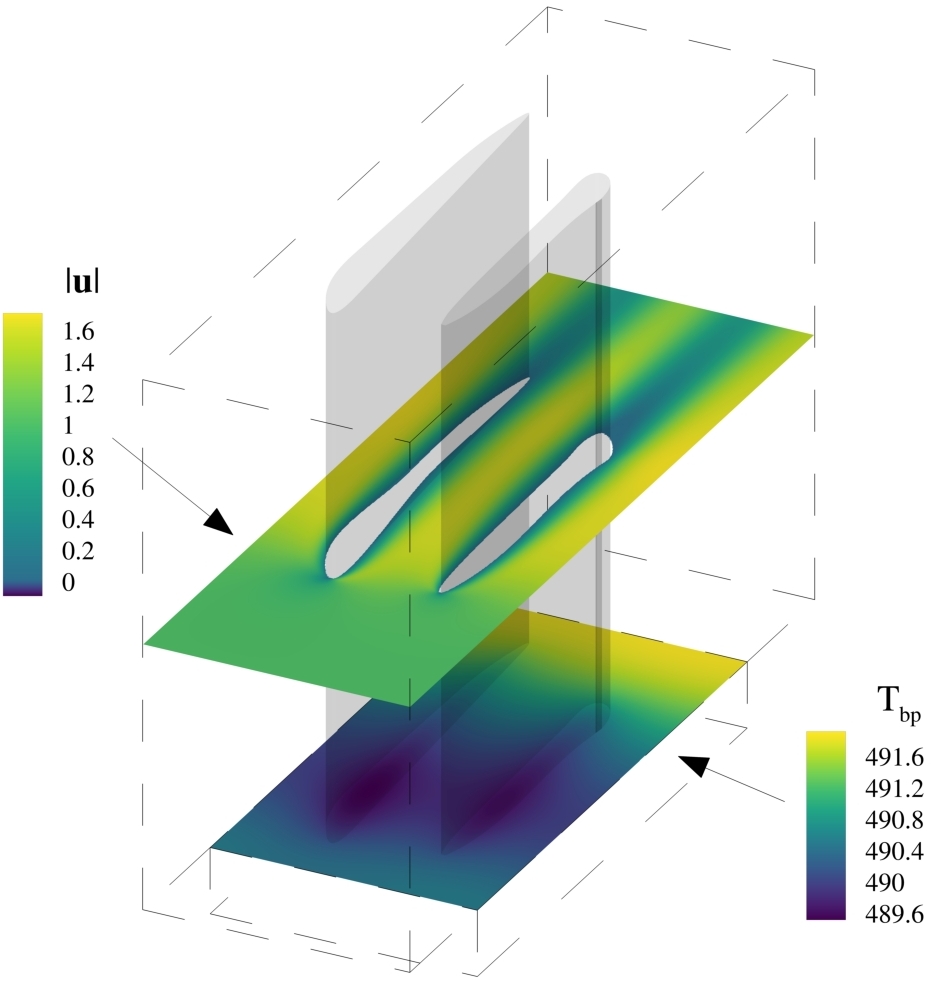}
        \captionsetup{labelformat=empty}
        \caption{(c)}
    \end{subfigure} 

%    \begin{subfigure}[b]{0.45\textwidth}
%        \includegraphics[width=\textwidth]{T_dist_number_2_cropped.pdf}
%        \captionsetup{labelformat=empty}
%        \caption{(c)}
%    \end{subfigure}
%    \hfill
%    \begin{subfigure}[b]{0.45\textwidth}
%        \includegraphics[width=\textwidth]{diff_T_number_2_cropped.pdf}
%        \captionsetup{labelformat=empty}
%        \caption{(d)}
%    \end{subfigure}

 %   \vspace{1em}  % Optional vertical spacing

 %       \begin{subfigure}[b]{0.45\textwidth}
 %       \includegraphics[width=\textwidth]{T_dist_number_3_cropped.pdf}
 %       \captionsetup{labelformat=empty}
 %       \caption{(e)}
 %   \end{subfigure}
 %   \hfill
 %   \begin{subfigure}[b]{0.45\textwidth}
 %       \includegraphics[width=\textwidth]{diff_T_number_3_cropped.pdf}
 %       \captionsetup{labelformat=empty}
 %       \caption{(f)}
 
 %   \end{subfigure}

    \caption{Velocity magnitude ($|u|$) and base plate temperature ($T_{bp}$) distributions for (a) rectangular, (b) air-foil, and (c) arbitrary shape fins. (a) base plate average temperature, $\overline{T}_{\Omega_{bp}}$, and pressure drop value, $\Delta\overline{p}$, for pseudo 3D approach are 508.62(K) and -1.096(Pa), respectively, and for 3D full scale simulation are 517.75(K) and -1.348(Pa) (b) $\overline{T}_{\Omega_{bp}}$, and $\Delta\overline{p}$ for pseudo 3D approach are 549.86(K) and -0.71(Pa), respectively, and for 3D full scale simulation are 558.06(K) and -0.917(Pa) (c) $\overline{T}_{\Omega_{bp}}$, and $\Delta\overline{p}$ for pseudo 3D approach are 490.49(K) and -0.931(Pa), respectively, and for 3D full scale simulation are 497.12(K) and -1.148(Pa).}
    \label{fig:validation_combined}
\end{figure}

\begin{figure}[H]
    \centering

    % Top row: (a) and (b)
    \begin{subfigure}[b]{0.45\textwidth}
        \includegraphics[width=\textwidth]{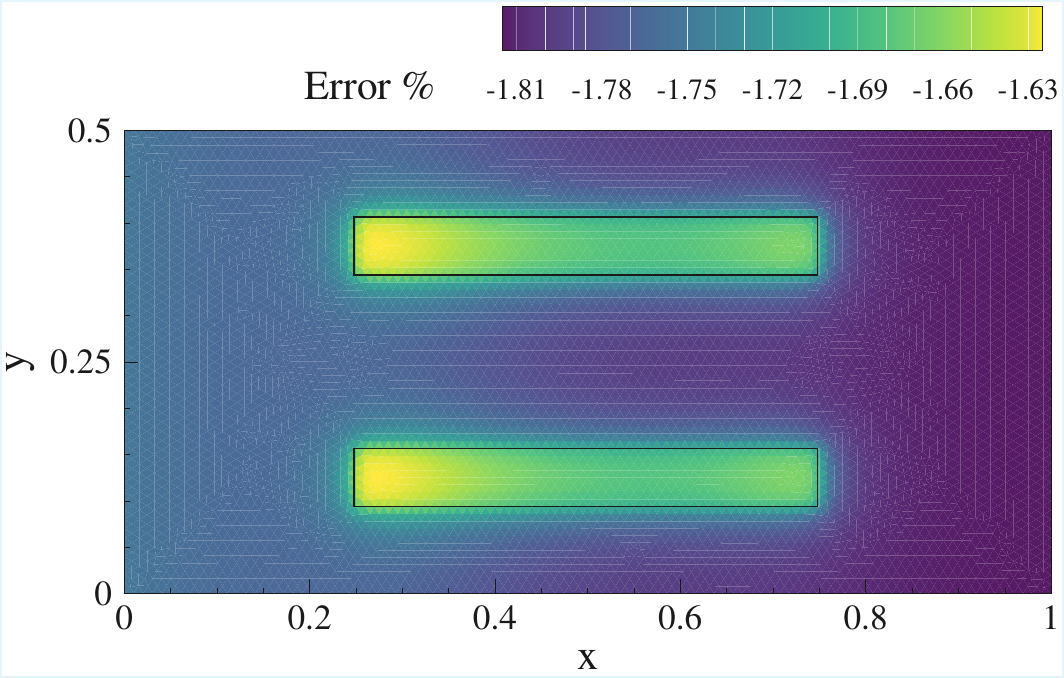}
        %{u_full_3d_cropped_1.pdf}
        \captionsetup{labelformat=empty}
        \caption{(a)}
    \end{subfigure}
    \hfill
    \begin{subfigure}[b]{0.48\textwidth}
        \includegraphics[width=\textwidth]{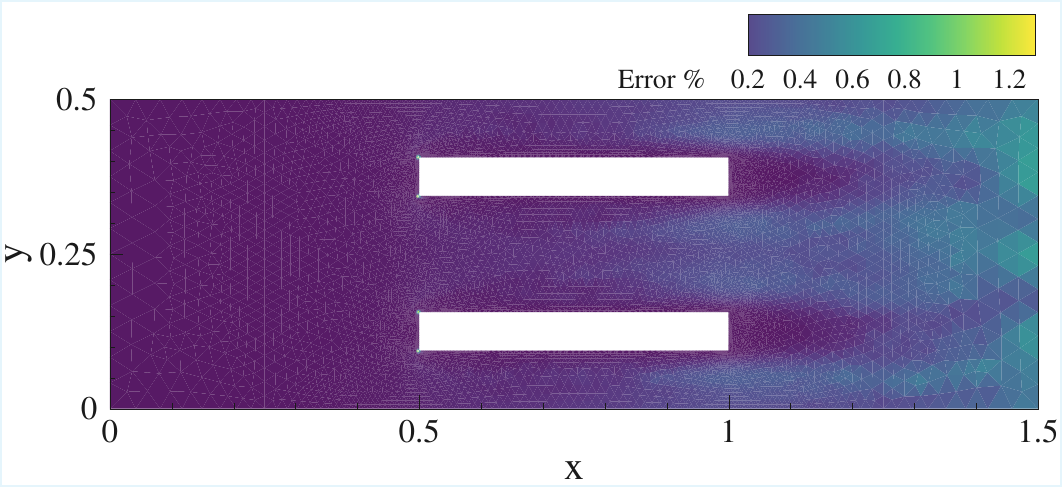}
        \captionsetup{labelformat=empty}
        \caption{(b)}
    \end{subfigure}

    \vspace{1em}  % Optional vertical spacing

    % Top row: (a) and (b)
    \begin{subfigure}[b]{0.45\textwidth}
        \includegraphics[width=\textwidth]{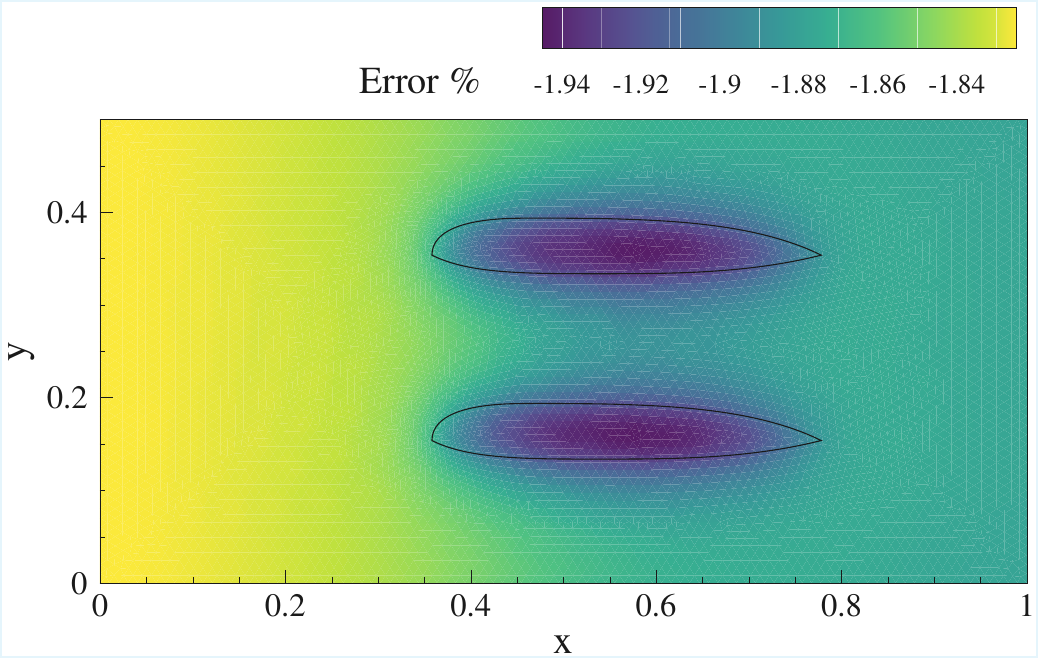}
        %{u_full_3d_cropped_2.pdf}
        \captionsetup{labelformat=empty}
        \caption{(c)}
    \end{subfigure}
    \hfill
    \begin{subfigure}[b]{0.48\textwidth}
        \includegraphics[width=\textwidth]{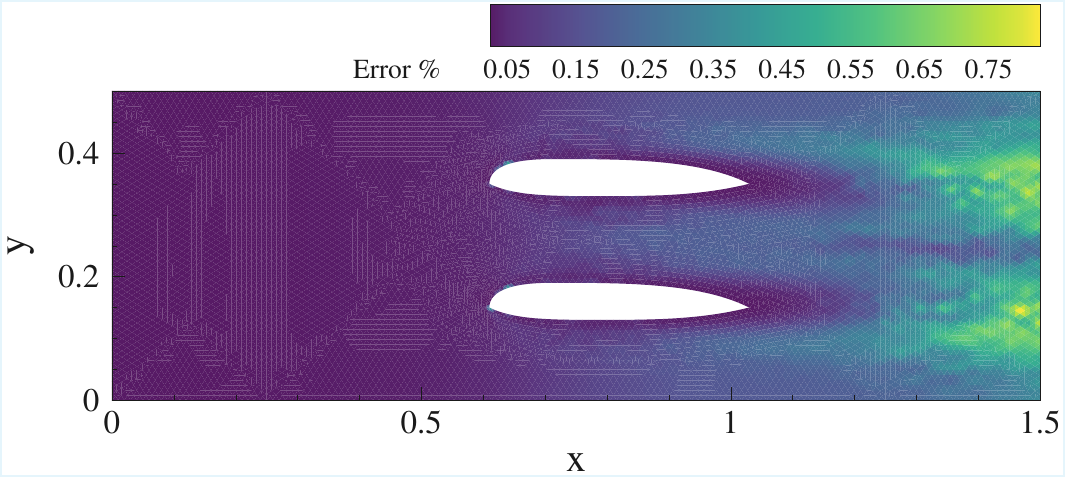}
        \captionsetup{labelformat=empty}
        \caption{(d)}
    \end{subfigure}

    \vspace{1em}  % Optional vertical spacing
    % Top row: (a) and (b)
    \begin{subfigure}[b]{0.45\textwidth}
        \includegraphics[width=\textwidth]{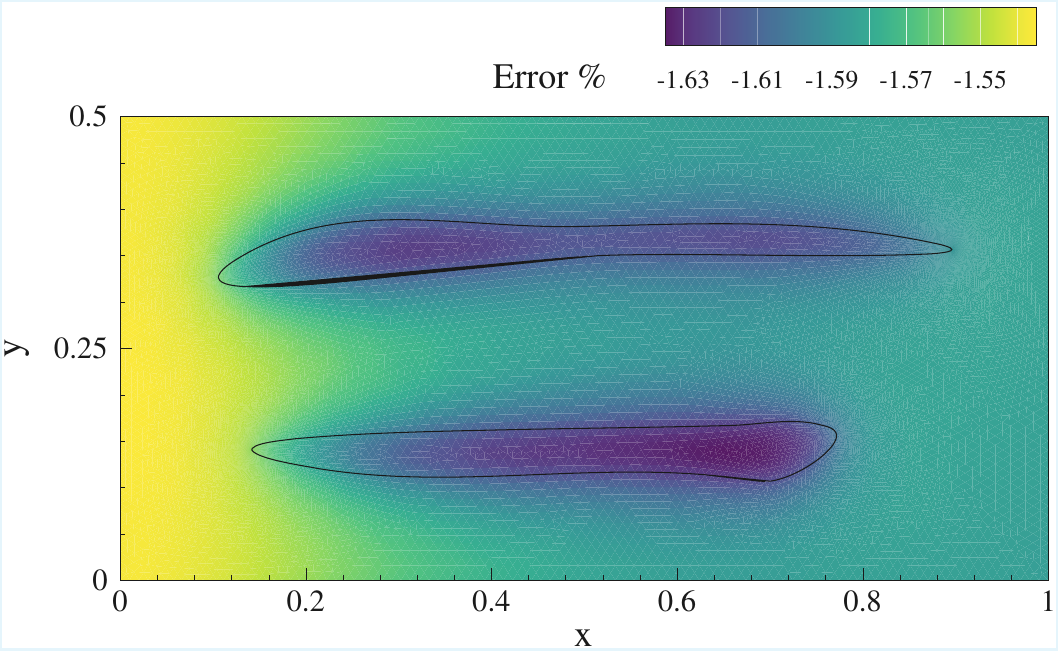}
        %{u_full_3d_cropped_3.pdf}
        \captionsetup{labelformat=empty}
        \caption{(e)}
    \end{subfigure}
    \hfill
    \begin{subfigure}[b]{0.48\textwidth}
        \includegraphics[width=\textwidth]{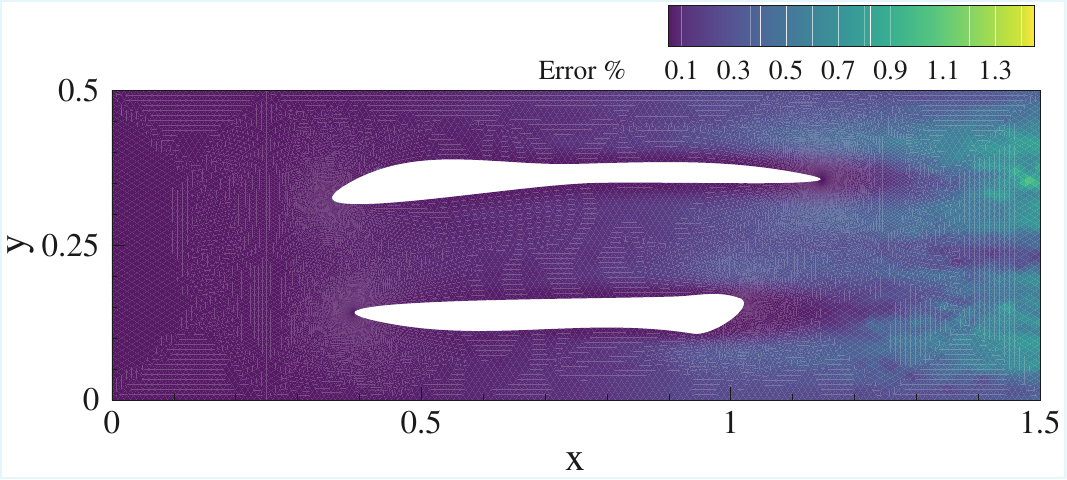}
        \captionsetup{labelformat=empty}
        \caption{(f)}
    \end{subfigure}

    \vspace{1em}  % Optional vertical spacing

    \caption{Temperature and velocity magnitude error distribution for the different cases shown in figure \ref{fig:validation_combined}. The temperature error distributions shown in panels (a), (c), and (e) are computed using the relative error formula, $Error= (T_{bp}-T_{3D})/T_{3D}$. For the velocity field error distributions in panels (b), (d), and (f), the normalized mean-square error is employed, $Error= (|{\mathbf{u}}|_{\text{pseudo-3D}} - |\mathbf{u}|_{\text{3D}})^{2} / |\mathbf{u}|^2_{\text{ave,3D}}$.}
    \label{fig:error_validation_combined}
\end{figure}

We begin presenting the results through comparisons between the full 3D simulations and the pseudo-3D approach, as shown in Figures \ref{fig:validation_combined} and \ref{fig:error_validation_combined}. Figure \ref{fig:validation_combined}  shows the velocity magnitude ($|u|$) and base plate temperature ($T_{bp}$) distribution for three different cases of a rectangular, an airfoil, and an arbitrary-shaped fin geometries. In all three cases, the pseudo-3D method properly captures the flow pattern and temperature gradients along the channel. The velocity magnitude is higher between the fins due to the reduced cross-sectional area. Correspondingly, the base plate temperature distribution reflects the cooling effect of the attached fins, with distinct low-temperature zones formed near the areas covered by the fins. The comparison of the base plate average temperatures and pressure losses against the full-scale 3D simulation, provided in Figure \ref{fig:validation_combined} caption, reveals that the pseudo-3D approach predicts values very close to the full 3D simulations, with only small deviations. The pressure loss values show a maximum deviation of less than 0.2 Pa between the pseudo approach and the full 3D simulation for all cases, which arises due to the influence of the top and bottom surfaces of the 3D-channel, which are neglected in the pseudo 3D approach.\\ Figure \ref{fig:error_validation_combined} quantifies the local error between the pseudo-3D approach and 3D results, by plotting the temperature error distribution on the left side (sub-figures a, c, e) and displaying the velocity field errors on the right side (sub-figures b, d, f). To quantify the temperature distribution error, the absolute relative error formula, $Error= (T_{bp}-T_{3D})/T_{3D}$, is considered, while for the velocity field, where the local velocity approaches zero at the stagnant points, the normalized mean square error, $Error= (|{\mathbf{u}}|_{\text{pseudo-3D}} - |\mathbf{u}|_{\text{3D}})^{2} / |\mathbf{u}|^2_{\text{ave,3D}}$, is used. For all three geometries, the temperature error remains within a small range (below ~2\%), which confirms that the values obtained for the parameters $h_f$ and $h_s$ are appropriate. It is also visible that the velocity error is mostly localized downstream of the channels with high velocity gradients, showing a maximum deviation of less than 1.5\%. Overall, the comparison indicates that the pseudo-3D approach properly captures the velocity, pressure, and temperature fields, thereby offering a close approximation to the full 3D solution. 

%%%

\begin{figure}[H] % Requires \usepackage{float}
    \centering
    \includegraphics[width=1.0\textwidth]{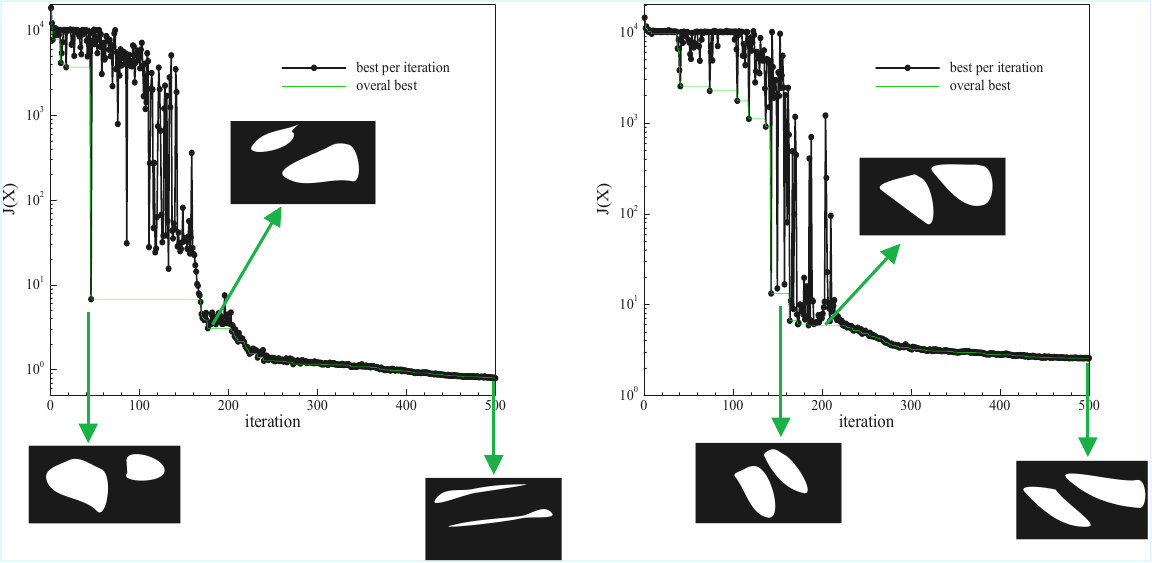}
    \caption{CMA-ES Convergence plot for $\overline{T}_\mathrm{cons}=500\mathrm{K}$ (left one) and $\overline{T}_\mathrm{cons}=475\mathrm{K}$ (right one). The final optimized value for \(\Delta p/\Delta p_{\mathrm{SF}}\) after 500 iterations is 0.552 and 0.828, respectively. The insets illustrate intermediate and final optimized geometries obtained during the optimization procedure. }
    \label{fig:convergence}
\end{figure}

Figure \ref{fig:convergence} represnts the convergence history of the CMA-ES algorithm for shape optimization under two thermal constraints: $\overline{T}_\mathrm{cons}=500K$ (left) and $\overline{T}_\mathrm{cons}=475K$ (right). The plots show the objective function, $J(X)$, versus iteration, including the best solution per iteration (black) and the overall best (green). The insets illustrate intermediate and final optimized geometries obtained during the optimization process. The pressure losses for straight fins, $\Delta{p}_\mathrm{SF}$, corresponding to the same surface average temperature, are 1.429 and 3.05 Pa for the 500K and 475K cases, respectively. This indicates that by using the optimized geometries, the pressure loss can be improved by almost 45 \% and 20 \%, respectively. In both cases, the variance of the candidate solutions is initially high, and the algorithm initially explores a wide range of different shapes, reflected in large oscillations of the cost function value. This is an expected behavior, as the algorithm samples widely to learn the range of feasible solutions. After approximately 150–200 iterations, the produced offspring converge toward stable shapes with significantly reduced cost function or equivalently pressure drop value. In this step, the algorithm mainly exploits promising regions, and progressively improves the fin shape. \\For the 500K temperature constraint (left), the optimization converges to an elongated channel-like geometry that promotes lower pressure losses while still satisfying the temperature constraint. For the stricter 475K case (right), the final design is more compact, balancing the need for a lower surface temperature with an increased pressure loss. 

\begin{figure}[H] % Requires \usepackage{float}
    \centering
    \includegraphics[width=1.0\textwidth]{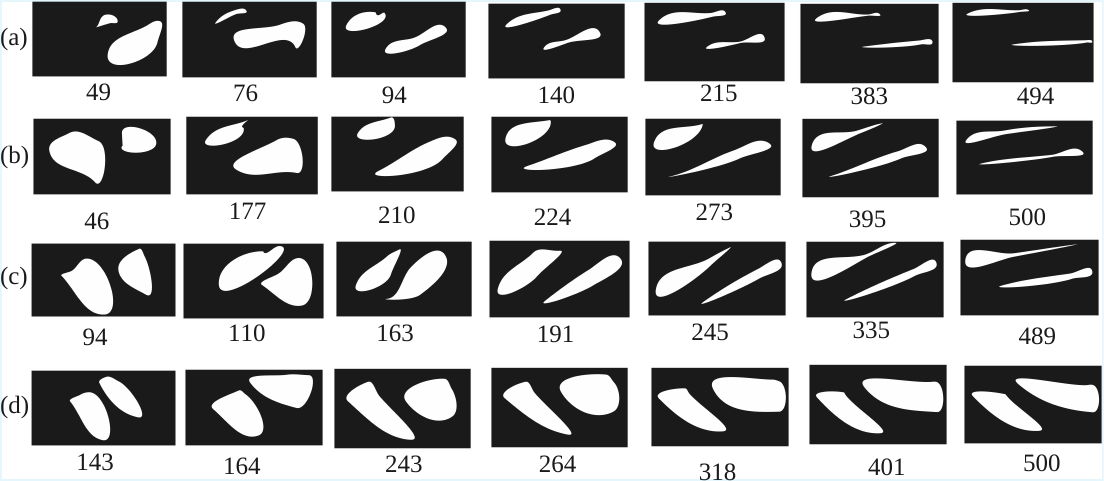}
    \caption{CMA-ES–driven shape evolution toward the optimized geometry. Shown are intermediate elites and the final design, which minimizes the pressure drop under the imposed constraint, (a) $\overline{T}_\mathrm{cons}=550\mathrm{K}$, (b) $\overline{T}_\mathrm{cons}=500\mathrm{K}$, (c) $\overline{T}_\mathrm{cons}=487.5\mathrm{K}$, (d) $\overline{T}_\mathrm{cons}=475\mathrm{K}$.  The optimized \(\Delta p/\Delta p_{\mathrm{SF}}\) values are 0.86, 0.55, 0.49, and 0.82,  respectively.}
    \label{fig:shape_evolution}
\end{figure}

Figure \ref{fig:shape_evolution} shows the shape evolution process obtained using the CMA-ES algorithm under different thermal constraints. Each row corresponds to a distinct average temperature constraint and the images represent the fin shapes (white) within the flow domain (black) at different optimization iterations (labeled below each snapshot). The shape evolution results indicate that the algorithm prioritizes pressure reduction under less restrictive thermal limits (i.e., $\overline{T}_\mathrm{cons}=550 \mathrm{K}, 500 \mathrm{K}$), leading to more slender and streamlined shapes. In contrast, the shape evolution algorithm emphasizes thermal performance and hot spot prevention under more stringent thermal constraints (i.e., $\overline{T}_\mathrm{cons}=475 \mathrm{K}$), resulting in bulkier fins at the expense of increased pressure losses. The balance between exploration and exploitation during the progress is also visible from the figure. At the early stages of the shape evolution process, where the algorithm mainly focuses on exploration, the shapes are more diverse, whereas by approaching the later stages, exploitation overcomes, resulting in shapes that are closer to optimal designs. The ratios of the pressure drop of the optimized designs to that of the straight fins, satisfying the same thermal constraints of 550 K, 500 K, 487.5 K, and 475 K, are 0.86, 0.55, 0.49, and 0.82, respectively. Therefore, the results reveal that the optimized shapes reduce the pressure drop penalty by up to almost 50 percent, depending on the imposed thermal constraint. 

\begin{figure}[H] % Requires \usepackage{float}
    \centering
    \includegraphics[width=0.90\textwidth]{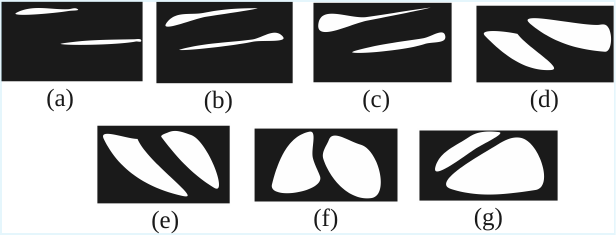}
    \caption{Final optimum designs obtained after 500 iterations, satisfying (a) $\overline{T}_\mathrm{cons}=550\mathrm{K}$, (b) $\overline{T}_\mathrm{cons}=500\mathrm{K}$, (c) $\overline{T}_\mathrm{cons}=487.5\mathrm{K}$, (d) $\overline{T}_\mathrm{cons}=475\mathrm{K}$, (e) $\overline{T}_\mathrm{cons}=450\mathrm{K}$, (f) $\overline{T}_\mathrm{cons}=425\mathrm{K}$, and (g) $\overline{T}_\mathrm{cons}=400\mathrm{K}$.   }
    \label{fig:shape_evolution_Finals}
\end{figure}

Figure \ref{fig:shape_evolution_Finals} represents the optimized designs after 500 iterations for different temperature constraints. A clear trade-off between pressure drop and thermal performance is observed. Under less stringent thermal constraints (parts a-c), the algorithm leads to slender, streamlined shapes, whereas strict thermal limits (parts d-g) drive shape thickening. A comparison between the average surface temperature and pressure drop of the final optimized designs, shown in Figure \ref{fig:shape_evolution_Finals}, for the pseudo-3D and full-scale 3D approaches is made in Table \ref{tab:pseudo_vs_3D}. As can be seen, there is a good agreement between the pseudo-3D and full 3D results. The minor difference between the pressure drop values roots in the fact that the effects of the top and bottom surfaces in the pseudo-3D approach are neglected. Despite the small difference between the two approaches, the simulation time required for the pseudo-3D method is significantly lower than that of the full 3D simulation. This makes the pseudo-3D approach far more practical for optimization problems, where a large number of iterations are needed to reach an optimal solution.

\begin{table}[h!]
\centering
\caption{Comparison of evaluated average base plate temperature and pressure drop values between pseudo-3D and full 3D simulations under various temperature constraints.}
\begin{tabular}{|c|c|c|c|c|}
\hline
$\overline{T}_{\mathrm{cons}}$ [K] & \multicolumn{2}{c|}{Pseudo-3D} & \multicolumn{2}{c|}{Full 3D} \\
\hline
 & $\overline{T}_{\mathrm{ave}}$ [K] & $\Delta p$ [Pa] & $\overline{T}_{\mathrm{ave}}$ [K] & $\Delta p$ [Pa] \\
\hline
550   & 549.92 & 0.47288 & 556.93 & 0.666 \\
500   & 499.73 & 0.78950 & 508.69 & 1.0013 \\
487.5 & 487.35 & 0.96700 & 494.33 & 1.1987 \\
475   & 474.91 & 2.52880 & 478.66 & 2.8147 \\
450   & 449.89 & 5.64730 & 450.42 & 6.006 \\
425   & 424.86 & 24.2700 & 425.13 & 24.349 \\
400   & 399.99 & 38.0570 & 400.78 & 38.394 \\
\hline
\end{tabular}
\label{tab:pseudo_vs_3D}
\end{table}

\begin{figure}[H]
    \centering

    % Top row: (a) and (b)
    \begin{subfigure}[b]{0.48\textwidth}
        \includegraphics[width=\textwidth]{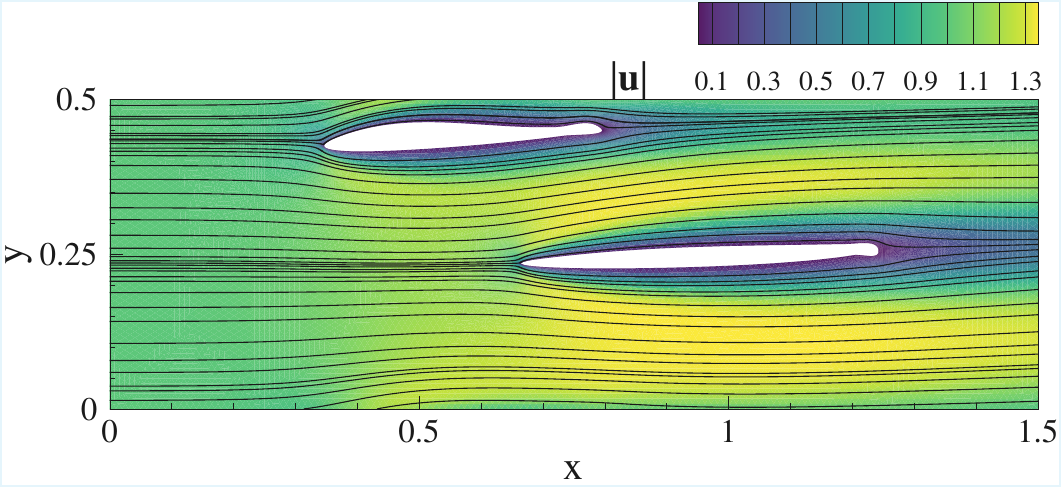}
        \captionsetup{labelformat=empty}
        \caption{(a)}
    \end{subfigure}
    \hfill
    \begin{subfigure}[b]{0.48\textwidth}
        \includegraphics[width=\textwidth]{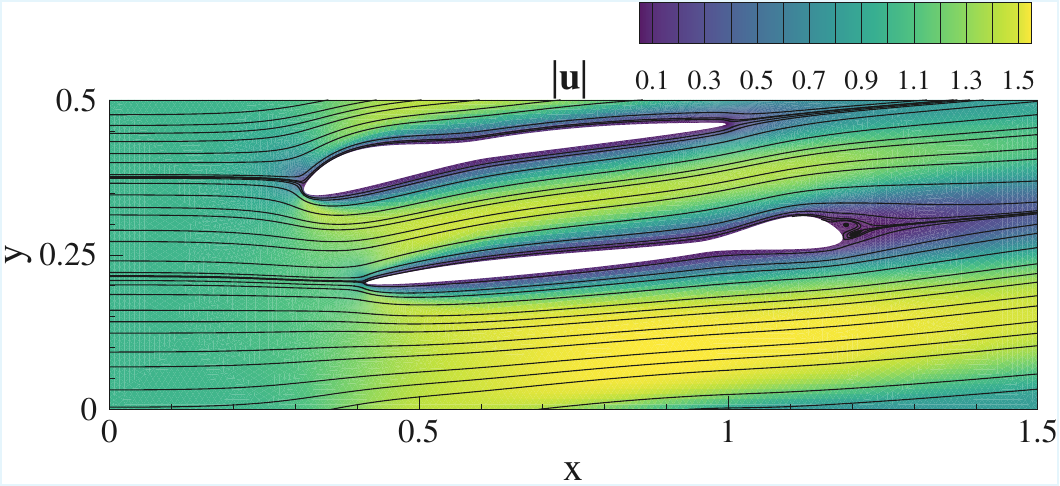}
        \captionsetup{labelformat=empty}
        \caption{(b)}
    \end{subfigure}

    \vspace{1em}  % Optional vertical spacing
    % Top row: (a) and (b)
    \begin{subfigure}[b]{0.48\textwidth}
        \includegraphics[width=\textwidth]{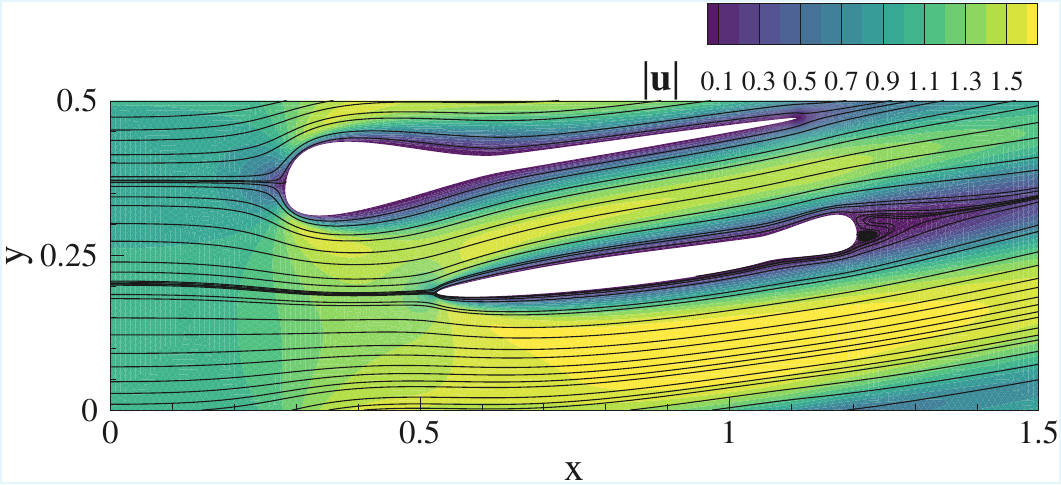}
        \captionsetup{labelformat=empty}
        \caption{(c)}
    \end{subfigure}
    \hfill
    \begin{subfigure}[b]{0.48\textwidth}
        \includegraphics[width=\textwidth]{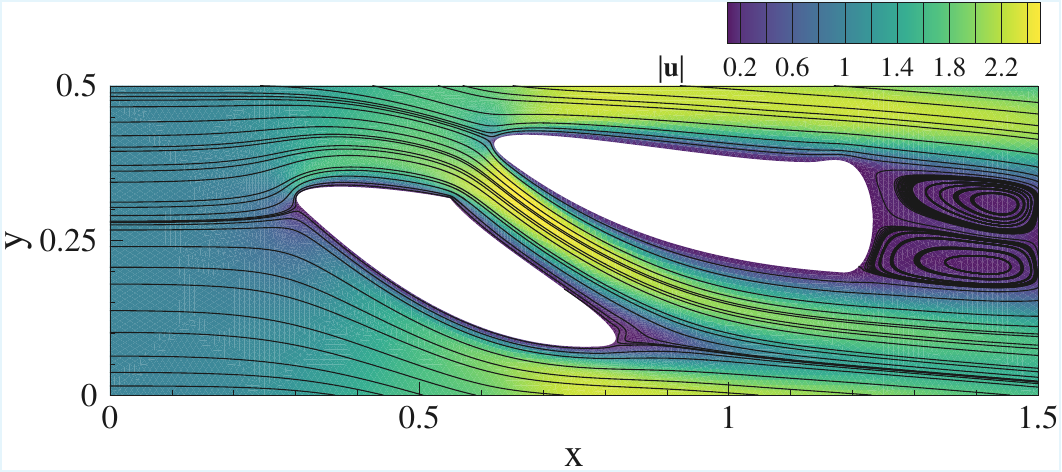}
        \captionsetup{labelformat=empty}
        \caption{(d)}
    \end{subfigure}
    \vspace{1em}  % Optional vertical spacing
        % Top row: (a) and (b)
    \begin{subfigure}[b]{0.48\textwidth}
        \includegraphics[width=\textwidth]{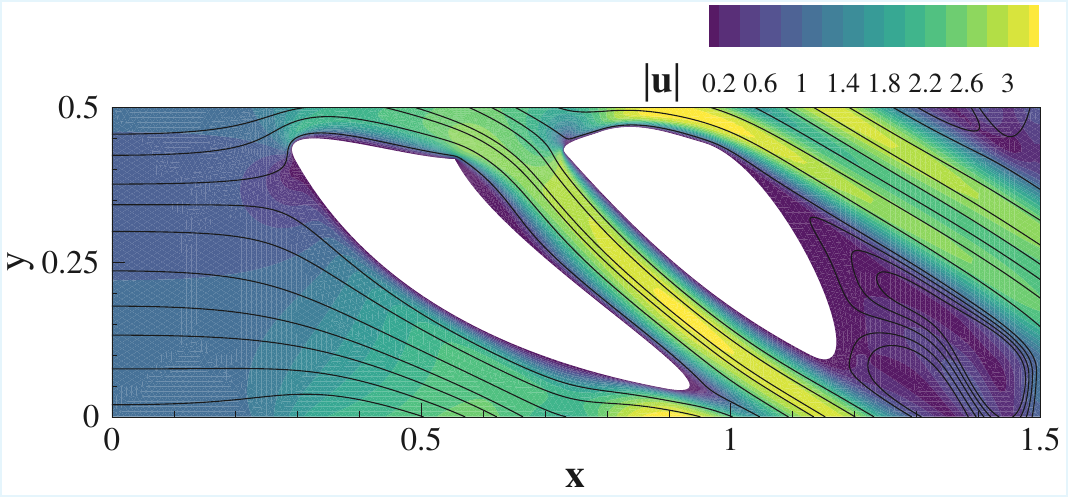} 
        %{velocity_field_T450_cropped.pdf}
        \captionsetup{labelformat=empty}
        \caption{(e)}
    \end{subfigure}
    \hfill
    \begin{subfigure}[b]{0.48\textwidth}
        \includegraphics[width=\textwidth]{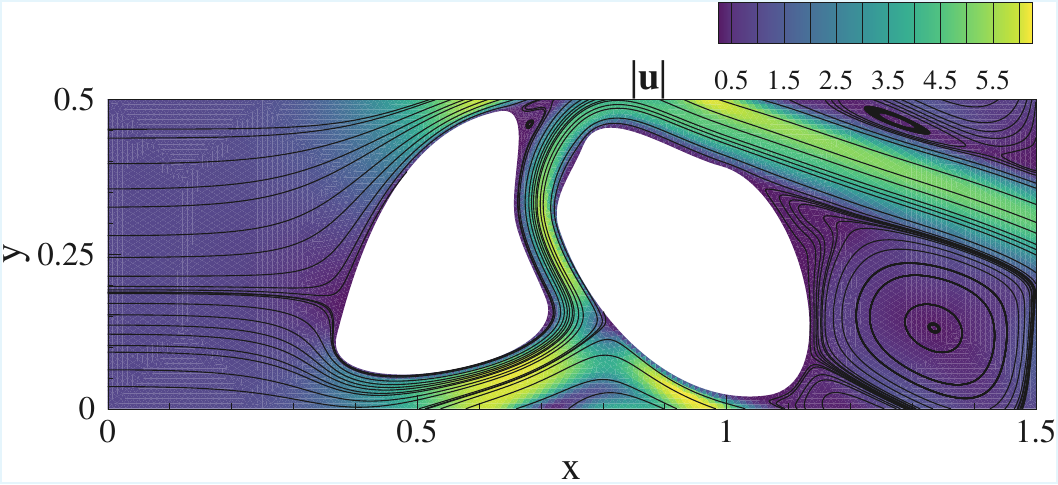}
        \captionsetup{labelformat=empty}
        \caption{(f)}
    \end{subfigure}
    % Bottom row: (c)
    \begin{subfigure}[b]{0.48\textwidth}
        \includegraphics[width=\textwidth]{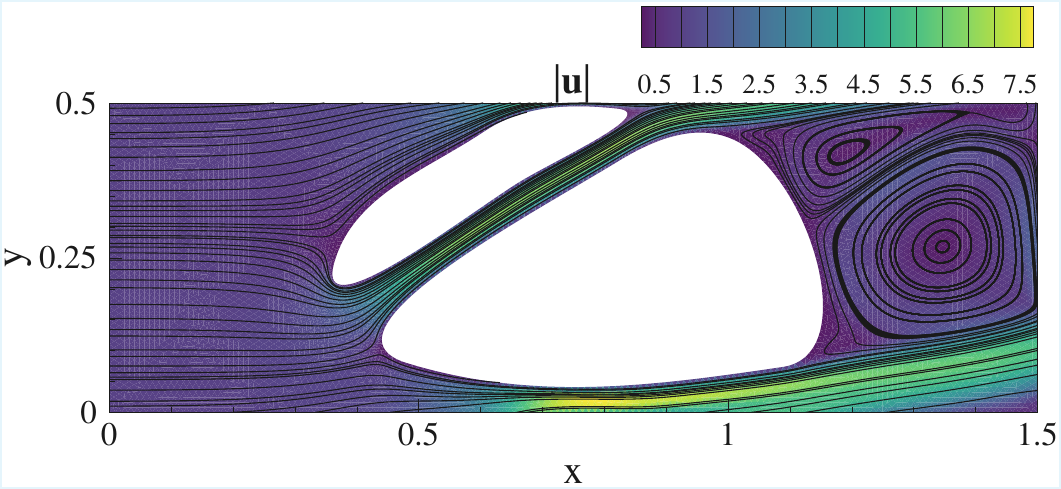}
        \captionsetup{labelformat=empty}
        \caption{(g)}
    \end{subfigure}

    \caption{Velocity magnitude distribution along with the streamlines around the final optimized designs shown in figure \ref{fig:shape_evolution_Finals}.}
    \label{fig:validation_combined_....}
\end{figure}

Figure \ref{fig:validation_combined_....} visualizes the flow behavior around the optimized fin designs, shown in Figure \ref{fig:shape_evolution_Finals}, as characterized by velocity magnitude and streamline contours. It is clear that the slender fin designs under less stringent temperature constraints (parts a–c) offer smooth and relatively undisturbed streamlines. This shape configuration leads to less viscous losses, resulting in lower pressure drop penalties. However, the more intrusive fin shapes obtained under more strict thermal limits (parts d–g) lead to the emergence of flow separation zones, recirculation regions, and vortices, particularly in the downstream areas of the fin structures. These regions intensify form drag and consequently elevated pressure losses. In addition to the emergence of form drag for designs with bulkier topologies, frictional drag also plays an important role when the flow pattern becomes more constricted in the narrow flow passages with high velocity gradients (like in part g).  The increment of both frictional and form drag contributes to more pressure losses as previously discussed.

\section{Conclusion}
In summary, an efficient and scalable multi-fidelity shape evolution algorithm was introduced to optimize the heat transfer features of heat sink devices under various thermal limitations. The algorithm employs Bézier curves, enabling flexible and smooth geometric representations of fin structures. The proposed shape optimization algorithm significantly reduces the computational cost by using a pseudo 3D approach, which simplifies the 3D problem into two thermally coupled 2D problems: (i) a 2D convection–conduction model representing heat transfer within the thermofluid layer across the fins, and (ii) a 2D conduction model for the base plate, accounting for heat spreading and heat exchange with the thermofluid layer. This decomposition preserves the essential 3D thermal physics while enabling rapid and computationally efficient shape optimization. The heat transfer coefficients between these two surfaces are calibrated using full 3D simulations and validated across different fin arrangements. The pseudo 3D approach makes it feasible to change the designs iteratively within the optimization algorithm while enabling rapid exploration of complex design spaces with good accuracy in estimating surface-average temperature and pressure drop. The shape-evolving procedure, as well as the implementation of the optimization method, is described and discussed in detail. We showed that the optimized designs offer up to almost 50\% improvement in pressure loss compared to conventional straight fins while satisfying prescribed surface-average temperature limits. The proposed shape optimization framework demonstrates a clear trade-off between thermal performance and hydraulic losses and offers a scalable, computationally efficient strategy for high-fidelity thermal optimization in future engineering applications.

\section*{Conflicts of interest}
There are no conflicts to declare.
%\nocite{*}
\bibliographystyle{unsrt}
\bibliography{references.bib}
%\bibliography{aipsamp}% Produces the bibliography via BibTeX.

\end{document}